\documentclass[12pt,a4paper]{article}   
\usepackage{graphicx}
\usepackage{amsmath,amsfonts,amssymb}

\title{Exact thermodynamics and Luttinger liquid properties
of the integrable $t-J$~model}
\author{G.~J\"uttner\thanks{e-mail: gj@thp.uni-koeln.de},        
        A. Kl\"umper\thanks{e-mail: kluemper@thp.uni-koeln.de},
        J. Suzuki\thanks{e-mail: js@thp.uni-koeln.de, 
        Permanent address: Institute of Physics, 
                           University of Tokyo at Komaba}\\
        \parbox{0.9\textwidth}{
        {\em
        \begin{center}
        Universit\"at zu K\"oln  \\                
        Institut f\"ur Theoretische Physik \\     
        Z\"ulpicher Str. 77  \\                   
        D-50937, Germany
        \end{center}
        }}  
       }    

\date{October 1996}                          
\begin{document}

\maketitle
\begin{abstract}
  A Trotter-Suzuki mapping is used to calculate the finite-temperature
  properties of the  one-dimensional supersymmetric $t-J$ model.  This
  approach allows for the exact calculation of various thermodynamical
  properties by means of the  quantum transfer matrix (QTM).  The free
  energy and other interesting   quantities are obtained such   as the
  specific heat  and compressibility.  For  the largest  eigenvalue of
  the QTM leading  to  the free energy   a set of just two  non-linear
  integral  equations  is  presented.   These   equations are  studied
  analytically and numerically   for different particle  densities and
  temperatures.  The structure  of  the specific heat is  discussed in
  terms of the elementary charge as well as spin excitations.  Special
  emphasis  is  placed on the study  of  the  low-temperature behavior
  confirming    scaling predictions   by  conformal  field  theory and
  Luttinger liquid theory. To our knowledge this is the first complete
  investigation of a strongly correlated  electron system on a lattice
  at finite temperature. 
\end{abstract}

\clearpage
\section{Introduction}
\label{sec:intro}

Strongly correlated   electron  systems  have  attracted  considerable
interest   in  recent years   in view   of mechanisms  for  high-$T_c$
superconductivity  \cite{And87,ZhaRi88}.    The $t-J$ model  arises in
various contexts and  represents one of  the most fundamental systems.
The model describes  the nearest-neighbor  hopping of  electrons  with
spin-exchange interaction.  The effect  of a strong  repulsive on-site
Coulomb interaction is modeled by the restriction of the Hilbert space
to states without doubly  occupied lattice sites.  The one-dimensional
Hamiltonian reads 
\begin{equation}
  {\cal H}=-t\sum_{j,\sigma}{\cal P}
                     (c^\dagger_{j,\sigma}c_{j+1,\sigma}+
                      c^\dagger_{j+1,\sigma}c_{j,\sigma})
                     {\cal P} 
    +J\sum_j(S_j S_{j+1}-n_j n_{j+1}/4),
  \label{hamilton}  
\end{equation}
where                               the                      projector
\mbox{${\cal{P}}=\prod_j(1-n_{j\uparrow}n_{j\downarrow})$}     ensures
that double occupancies of sites are forbidden.  At the supersymmetric
point $2t=J$ the system was shown to be integrable \cite{Suth75,Schlott87}
by the well-known  Bethe ansatz \cite{Bethe31,Yang67}.  The ground state
and excitation   spectrum   were   investigated  \cite{BarBlaOg91} and
critical  exponents  calculated  by finite-size  scaling and conformal
field theory  studies \cite{KawaYa90,BarKlu95}.  Here we  will discuss
the model at finite temperatures. 

According    to    the   seminal  work   \cite{YYang69,Gaudin71,Tak71}
thermodynamical properties of general integrable systems are described
by an infinite   set  of  coupled integral equations   reflecting  the
existence of  infinitely many  different  rapidity patterns.  For  the
$t-J$    model    such  integral   equations    were    formulated  in
\cite{Will92,Schlott92,Schlott96}.   The method  consists of a  direct
evaluation  of the  partition  function  by taking  into   account all
excited states of the Hamiltonian.  The excitations are derived on the
basis of the so-called string conjecture which describes the solutions
to the Bethe ansatz equations (for details see \cite{FoKa93}).  In the
evaluation of finite-size  quantities at zero temperature the validity
of        the  string      conjecture    is        controversial
\cite{Woy82,BabdeVV83,KluZitt89,Vla84,EssKoSch92:JPA,JuDo93}.          
Sometimes, a simplistic  application  is  known to lead  to  erroneous
results  \cite{AlMar88}.   On the  other hand,  the string approach to
thermodynamics apparently yields exact  equations.  (See for  instance
\cite{Klu92}  where the  traditional   thermodynamics of the   quantum
$RSOS$ chains were recovered  by an independent approach utilizing the
fusion hierarchy analysis.)  For a nice  illustration of the successes
of the traditional approach  to thermodynamic properties the reader is
referred to the review \cite{TsvWieg83} on the Kondo problem. 

The general problem to be faced within the traditional string approach
to thermodynamics is often the necessity  to deal with infinitely many
integral equations.  In view of practical computations this requires a
truncation   scheme which is  difficult  to  control.   Also, and more
fundamentally, the calculation    of quantities other than   the  free
energy, for instance correlation lengths at finite temperature, is not
possible in the traditional approach. 

Our  approach takes  a  different route  to  overcome  these problems. 
Quite     generally,  $d$-dimensional  quantum     systems   at finite
temperatures can be mapped onto classical systems on $d+1$-dimensional
inhomogeneous  lattices by a  Trotter-Suzuki mapping \cite{Suz85}.  We
therefore  employ a convenient  mapping to a two-dimensional classical
model.  The classical   system corresponding to   the integrable $t-J$
chain       \cite{Bar94}        is     the     Perk-Schultz      model
\cite{PerkSchulz81,Schulz81,Schulz83,deVeLo91}  which  is a well-known
multi-component   generalization   of  the    six-vertex   model.  The
interesting thermodynamical quantities are expressed by eigenvalues of
an  appropriately  defined  operator, the   so-called quantum transfer
matrix (QTM).   The largest eigenvalue,  as usual, directly yields the
free  energy.  The further  knowledge  of the next-largest eigenvalues
provides the   correlation lengths.    This represents  the  important
advantage  \cite{Suz85,SuIn87,Koma87,Tsune91,SuAkWa90} compared to the
traditional thermodynamical Bethe  ansatz requiring all eigenvalues of
the Hamiltonian. 

In    addition to our general  strategy    we will incorporate another
important ingredient.  The most common structure of solvable models is
the  existence of  a family   of   commuting matrices  comprising  the
physical operator    (Hamiltonian or transfer   matrix).  We therefore
utilize a  mapping to a classical  system where the  QTM is apparently
embedded   in  such a family.    While  the general  Trotter arguments
respect the first conserved    quantity only, our choice implies   the
existence of infinitely many  conserved quantities.  This  assures the
integrability of the  QTM from the  very beginning.  The  new approach
has   been         applied     to   several       quantum      systems
\cite{SuAkWa90,SuNaWa92,Tak91,Klu92,Klu93,DeVe92,Mizuta95,DeVe95}    
which   essentially reduce  to  scalar field    theories.  
In particular a finite set of integral equations  has been derived for
the   spin    $1/2$  Heisenberg   chain    and  related    models   in
\cite{Klu92,DeVe92,Klu93,DeVe95}.  The characteristic feature of these
equations is the compact formulation originating from the very closely
related  calculation  of finite-size eigenvalues  of transfer matrices
\cite{KluBaPe91}. 
The investigations  of  highly
correlated   electron systems, such as   the Hubbard  model, have just
started in \cite{KluBa95} along the novel strategy.  

Here we adopt the most
sophisticated  method \cite{Klu93}.  The   QTM is of  the ``staggered"
type and is labeled by two spectral parameters.  The dependence on the
first parameter leads  to the commuting family  and thereby playing an
essential  role    in  the   diagonalization  procedure.   The  second
parameter,   on  the other  side,  intertwines  the finite temperature
system with the finite-size geometry  of the underlying lattice.  This
spectral parameter   represents  a  spatial anisotropy   of  Boltzmann
weights in  the  two-dimensional model.   Thus  our procedure  can  be
regarded as a  finite temperature extension  of the usual  Hamiltonian
limit.   However, we  must  be careful in taking  the  limit to obtain
finite temperature properties: a fine tuning between the parameter and
the  system size is  necessary. This is the price   to pay for dealing
with non-vanishing temperatures.  We will overcome the last problem by
adopting  the method developed  in  the finite-size correction problem
\cite{KluBaPe91}.   
All  information of the   Bethe ansatz equations will be
transformed  into  {\it   finitely  many} coupled   nonlinear integral
equations, valid  for any system sizes.  The Trotter limit then can be
taken analytically.    The   resultant  equations  yield  the    exact
thermodynamical properties of the model. 

The  study of  the thermodynamics of   the $t-J$  model using  the QTM
method has  been  introduced in  a short  letter \cite{JuKlu96}.   The
present    paper is  devoted   to    a more comprehensive   analytical
investigation  of this  system.   This   work constitutes the    first
approach   allowing for   analytical   and   high  precision numerical
treatments of a multicomponent  Luttinger liquid.  In fact, we confirm
the general    Luttinger  liquid picture   of   interacting   fermions
analytically in the low-temperature  limit. Furthermore, in this limit
our equations reduce  to {\it linear}  integral equations thus  making
direct contact  with the dressed energy  formalism for the groundstate
investigation of the Hamiltonian \cite{BoIzKo86,IzKoRe89,KawaYa91}. 

The paper is organized as follows.  In Section~\ref{sec:qtm} we derive
the quantum transfer matrix based on the  Perk-Schulz model by keeping
the   integrability    structure   (see   also   \cite{KluWe96}).   In
Section~\ref{sec:bae} the eigenvalue equations of  the QTM are derived
by an algebraic Bethe ansatz.  In Section~\ref{sec:nle} the eigenvalue
equations  are transformed   into  non-linear integral  equations  and
general    results   for    physical    quantities    are  presented.  
Section~\ref{sec:analytical}  deals  with several limiting cases which
are studied  analytically.  Section~\ref{sec:conclusion} contains  the
conclusion  of this work.    Several more technical, however important
aspects of the investigation are deferred to two appendices. 

\section{Quantum transfer matrix}
\label{sec:qtm}

The (classical) Perk-Schultz model  is considered on a square  lattice
with periodic  boundary  conditions.   Each  bond of  the   lattice is
occupied by   a variable  taking   on values  \mbox{$1,\dots,q$}.  The
appropriate          non-zero    Boltzmann      weights         ${\cal
  R}_{\alpha\beta}^{\mu\nu}$ associated  with   a vertex configuration
$\alpha, \beta, \mu, \nu$ on the lower, upper, left, and right bond 
\begin{equation}
  \begin{split}
    {\cal R}_{\alpha\alpha}^{\alpha\alpha}(v)&= 
    \sinh(\eta+\epsilon_\alpha v)/\sinh\eta,\\
    {\cal R}_{\alpha\alpha}^{\mu\mu}(v)&= 
    \epsilon_{\alpha}\epsilon_{\mu}\sinh(v)/\sinh\eta,\\
    {\cal R}_{\mu\alpha}^{\alpha\mu}(v)&= 
    \exp({\rm sign}(\alpha-\mu) v),
  \end{split}
\label{Rmatrix}
\end{equation}
satisfy       the     Yang-Baxter     equation     \cite{PerkSchulz81}
($\alpha,\mu=1,\dots,q$). The  parameters $\epsilon_\alpha$  take only
discrete values  $\pm    1$ corresponding  to   bosonic or   fermionic
statistics of the  state  $|\alpha\rangle$.  The  Yang-Baxter equation
implies the  commutation  of  all  row-to-row  transfer  matrices  for
arbitrary spectral parameters   $u$,   $v$:   
\mbox{${\cal   T}(u){\cal  T}(v)={\cal T}(v){\cal T}(u)$} with
\begin{equation}
  {\cal T}^\beta_\alpha(v)=\sum_\mu\prod_{i=1}^N
  {\cal R}_{\alpha_i\beta_i}^{\mu_i\mu_{i+1}}(v),
\label{transfer}
\end{equation}
Consequently, the Hamiltonian  (\ref{hamilton})  with \mbox{$t=1$}  is 
obtained  as the logarithmic derivative at  $v=0$ (where ${\cal T}(0)$
reduces to the right-shift operator ${\cal T}_R$)
\begin{equation}
  {\cal H}=\frac{{\rm d}}{{\rm d}v}
  \ln {\cal T}(v){\Big|_{v=0}}=\sum_{i=1}^N{h}_i,
\label{logderi}
\end{equation}
and   turns out to  be    integrable.     For   the  isotropic   limit
\mbox{$\eta\to{0}$} (with a rescaling $v\to \eta v$) the non-vanishing
matrix elements of the local operators ${h}_i$ read 
\begin{equation*}
  ({h}_i)^{\alpha\alpha}_{\alpha\alpha}=\epsilon_\alpha,\quad
  ({h}_i)^{\alpha\mu}_{\mu\alpha}=\epsilon_{\alpha\mu},
\end{equation*}
yielding   the  $t-J$    model     in  the case       \mbox{$q=3$} and
\mbox{$\{\epsilon_1,\epsilon_2,\epsilon_3\}$}= \mbox{$\{++-\}$}.   For
$q=3$       and         \mbox{$\{\epsilon_1,\epsilon_2,\epsilon_3\}$}=
\mbox{$\{+++\}$}       one   obtains   the     Uimin-Sutherland  model
\cite{Uimin70,Suth75}     and          for       \mbox{$q=4$}      and
\mbox{$\{\epsilon_1,\epsilon_2,\epsilon_3,\epsilon_4\}$}=
\mbox{$\{++--\}$} the Essler-Korepin-Schoutens model \cite{EssKoSch92},
respectively.   For more details   about ``generalized $t-J$ systems''
see \cite{Suz92,Bar94}. 

For further algebraic manipulations we introduce the Boltzmann weights
$\overline{{\cal R}}$ and $\widetilde{{\cal R}}$ of two models related
to (\ref{Rmatrix}) by anticlockwise and clockwise $90^0$ rotations 
\begin{equation*}
  \overline{{\cal R}}_{\alpha\beta}^{\mu\nu}(v)=
  {{\cal R}}^{\alpha\beta}_{\nu\mu}(v), \qquad
  \widetilde{{\cal R}}_{\alpha\beta}^{\mu\nu}(v)=
  {{\cal R}}^{\beta\alpha}_{\mu\nu}(-v).
\end{equation*}
According to (\ref{logderi}) we can write
\begin{equation}
  {\cal T}(v)={\cal T}_R\, {\rm e}^{v{\cal H}+{\cal O}(v^2)}, \qquad
  {\overline{\cal T}}(v)={\cal T}_L\, {\rm e}^{v{\cal H}+{\cal O}(v^2)},
\label{qtm-exp}
\end{equation}
where ${\overline{\cal T}}$ is defined  in analogy to (\ref{transfer})
and ${\cal   T}_{R,L}$  are  the   right-  and left-shift   operators,
respectively.  By means of the substitution 
\begin{equation}
u=-\beta/N,
\label{u}
\end{equation}
where  $\beta$  denotes the inverse  temperature and  $N$   is a large
integer ``Trotter'' number we find 
\begin{equation}
  \Big({\cal T}(u)\,\overline{{\cal T}}(u)\Big)^{N/2}=
   {\rm e}^{-\beta{\cal H}+{\cal O}(1/N)}.
\label{qtm-element}
\end{equation}
The  partition function  of the quantum system 
\begin{equation}
  Z=\lim_{N\to\infty}{\rm Tr}
  \Big({\cal T}(u)\,\overline{{\cal T}}(u)\Big)^{N/2},
\label{Z}
\end{equation}
is identical to the partition function of an inhomogeneous Perk-Schulz
model with alternating rows    \cite{KluWe96}.  The technically   more
convenient column-to-column transfer matrix of such  a system is often
referred to as the quantum transfer matrix  (QTM).  Obviously, it is a
member of the following family of matrices 
\begin{equation}
  {\cal T}^{\cal QTM}(v)=\sum_\mu\prod_{i=1}^{N/2}
  {\cal R}_{\alpha_{2i-1}\,\beta_{2i-1}}^{\mu_{2i-1}\,\mu_{2i}}(v+u)
  \,\widetilde{{\cal R}}_{\alpha_{2i}\,\beta_{2i}}
            ^{\mu_{2i}\,\mu_{2i+1}}(v-u),
\label{qtm}  
\end{equation}
at \mbox{$v=0$}. The remarkable property of this family of matrices is
its  commutativity,   as the ${\cal   R}$   and $\widetilde{{\cal R}}$
operators  possess   the   same intertwiner  which     can be  derived
immediately from   the   Yang-Baxter equation solely   for  ${\cal R}$
\cite{Klu96}. 

The  free energy $f$  per  lattice site  is  obtained from the largest
eigenvalue of ${\cal T}^{\cal QTM}(0)$ 
\begin{equation}
  f=-\frac{1}{\beta}\lim_{L\to\infty}\ln\left(\frac{Z}{L}\right) 
   =-\frac{1}{\beta}\lim_{N\to\infty}\ln\Lambda_{{\rm max}}.
\label{free-energy}
\end{equation}
Furthermore, the    next-leading eigenvalues  yield   the  correlation
lengths $\xi$ of the static correlation functions 
\begin{equation}
  \frac{1}{\xi}=-\lim_{N\to\infty}
  \ln\left|\frac{\Lambda}{\Lambda_{{\rm max}}}\right|.
\end{equation}
Although leaving the evaluation  of the finite-temperature correlation
length  of the $t-J$ model  as a future  problem, we stress that there
is,  in principle,  no  obstacle in obtaining  it  in contrast to  the
traditional string approach. 

Concluding this section we  would like to  give  some comments on  the
treatment of  the  thermodynamics of Hamiltonian  (\ref{logderi})  for
different  particle densities and  magnetizations.  As  usual, this is
achieved most conveniently  by introducing appropriate external fields
(chemical potential $\mu$, magnetic field  $h$). This in turn modifies
(\ref{Z}) by an additional  factor  under the  trace typically of  the
kind $\exp(\beta\mu\sum_in_i+\beta h\sum_is_i)$, where $n_i$ and $s_i$
are   the  particle number   and spin   operators at   site $i$.   The
definition of the associated  quantum  transfer matrix (\ref{qtm})  is
modified  only   by a  boundary  term   \cite{Klu93} depending  on the
variable~$\mu_{N+1}$. 

\section{Algebraic Bethe ansatz}
\label{sec:bae}

As previously mentioned    the largest eigenvalue and the   next-leading
eigenvalues  of the quantum transfer  matrix yield the thermodynamical
quantities  of the quantum chain.   In  \cite{KluWe96}  these
eigenvalues were obtained by an application of the algebraic Bethe ansatz.   
The  monodromy matrix for this case is defined by an alternating product 
\begin{equation}
  {\cal L}_\lambda^{\lambda '}(v)=
  {\cal R}_{\alpha_{1}\,\beta_{1}}^{\lambda\,\mu_{2}}(v+u)\,
  \widetilde{{\cal R}}_{\alpha_{2}\,\beta_{2}}
     ^{\mu_{2}\,\mu_{3}}(v-u)\,\dots\,
  {\cal R}_{\alpha_{N-1}\,\beta_{N-1}}^{\mu^{N-1}\,\mu_{N}}(v+u)\,
  \widetilde{{\cal R}}_{\alpha_{N}\,\beta_{N}}
     ^{\mu_{N}\,\lambda'}(v-u),
\end{equation}
which is related to the  quantum transfer matrix (\ref{qtm}) by taking
the trace over the auxiliary space 
\begin{equation}
  {\cal T}^{\cal QTM}(v)={\rm Tr}_{\rm aux}{\cal L}(v)\equiv
  \sum_\lambda{\cal L}_\lambda^{\lambda}(v).
\end{equation}
The monodromy matrix ${\cal L}(v)$ satisfies the Yang-Baxter equation 
\begin{equation}
  {\cal R}^{\lambda\beta}_{\mu\nu}(v-w)\,
  {\cal L}^{\lambda '}_\beta(v)\,{\cal L}^{\mu '}_\nu(w)=
  {\cal L}^{\nu}_\mu(w)\,{\cal L}^{\beta}_\lambda(v)\,
  {\cal R}^{\beta\lambda '}_{\nu\mu '}(v-w).
\label{yang-baxter-rel}
\end{equation}
Using the ``N\'eel state'' 
\mbox{$|\Omega\rangle=|1,2,1,2,1,2,\dots\rangle$}
as a reference state we can construct  states $|\Psi\rangle$ by 
applications of creation   operators 
${\cal  L}^3_1(v_i)$ and  ${\cal L}^2_3(w_i)$ 
with  different   rapidities   $v_i$   and  $w_i$.    
These  states    are  shown   to   be eigenstates
\cite{KluWe96} of the quantum transfer matrix with eigenvalues 
\begin{equation}
  \Lambda(v)=\lambda_{-}(v)+\lambda_{+}(v)+\lambda_0(v),
  \label{evlambda}  
\end{equation}
where
\begin{equation}
  \begin{split}
    \lambda_{-}(v)&=\prod_{j}\frac{v-w_j+{\rm i}\,\epsilon_1}{v-w_j} 
\Big[(v-i u-{\rm i}\,\epsilon_1)(v+{\rm i}\,
    u)\Big]^{N/2} \, {\rm e}^{\beta\mu_1}, \\ 
    \lambda_{+}(v)&=\prod_{k}\frac{v-v_k-{\rm i}\,\epsilon_2}{v-v_k} 
\Big[(v+{\rm i}\, u+{\rm i}\,\epsilon_2)(v-{\rm i}\,
    u)\Big]^{N/2} \, {\rm e}^{\beta\mu_2}, \\ 
    \lambda_0(v)&=\prod_{j}\frac{v-w_j-{\rm i}\,\epsilon_3}{v-w_j}\,
                  (v+{\rm i}\, u)^{N/2} \,
    \prod_{k}\frac{v-v_k+{\rm i}\,\epsilon_3}{v-v_k} \, 
    (v-{\rm i}\, u)^{N/2} \,
    {\rm e}^{\beta\mu_3},
  \end{split}
  \label{evlambda_part}
\end{equation}
and for further convenience we have replaced $v$  by ${\rm i}\,v$. The
quantities $\beta$ and   $\mu_j$  denote the inverse   temperature and
external  fields coupling to the three  different quantum states.  The
explicit relation to the chemical   potential $\mu$ and the   external
magnetic     field        $h$     for       the      $t-J$       model
\mbox{$\{\epsilon_1,\epsilon_2,\epsilon_3\}$} =\mbox{$\{++-\}$} reads
\begin{equation}
\mu_1=\mu+h/2, \quad \mu_2=\mu-h/2, \quad \mu_3=0.
\end{equation}
The   particular set of  spectral   parameters \mbox{$\{w_j,v_k\}$} --
often  referred to as   roots or rapidities  --  is determined by  the
condition  that the `unwanted  terms' cancel in the eigenvalue problem
for $|\Psi\rangle$.  This   provides  an eigenvector with   eigenvalue
$\Lambda(v)$.  Using  the Yang-Baxter equation (\ref{yang-baxter-rel})
it turns out \cite{KluWe96}  that the parameters  \mbox{$\{w_j,v_k\}$}
have to  satisfy a system   of coupled equations.   These Bethe ansatz
equations read 
\begin{equation}
  \begin{split}
    \left(\frac{v_i+{\rm i}\, u+{\rm i}\,\epsilon_2}
               {v_i+{\rm i}\, u}\right)^{N/2}
    &=-\prod_{j}\frac{v_i-w_j-{\rm i}\,\epsilon_3}{v_i-w_j}
    \prod_{k}\frac{v_i-v_k+{\rm i}\,\epsilon_3}{v_i-v_k-{\rm i}\,
    \epsilon_2}\:
    {\rm e}^{\beta(\mu_3-\mu_2)}, \\
    \left(\frac{w_i-{\rm i}\, u-{\rm i}\,\epsilon_1}
               {w_i-{\rm i}\, u}\right)^{N/2}
    &=-\prod_{k}\frac{w_i-v_k+{\rm i}\,\epsilon_3}{w_i-v_k}
    \prod_{j}\frac{w_i-w_j-{\rm i}\,\epsilon_3}{w_i-w_j+{\rm i}\,
    \epsilon_1}\:
    {\rm e}^{\beta(\mu_3-\mu_1)}.
  \end{split}
\label{evbae}  
\end{equation}
It is convenient to  use an alternative approach  to the Bethe ansatz:
The   defining relations  for the    rapidities  -- the  Bethe  ansatz
equations -- are equivalent  to the analyticity  of $\Lambda(v)$ as  a
function of $v$, i.e.   the absence of poles  in (\ref{evlambda}).  In
this  sense the  denominators in  (\ref{evlambda}) require immediately
the Bethe ansatz equations (\ref{evbae}).   We like to point out  that
any   appropriate treatment   of the    eigenvalue equations rendering
$\Lambda(v)$ analytic may  be considered as  an implicit determination
of the unknown roots.  This idea is used in the next section in a very
essential way. 

Next, some general properties of the Bethe ansatz roots are discussed.
Consider               the       three   different               cases
\mbox{$\{\epsilon_1,\epsilon_2,\epsilon_3\}$}    =\mbox{$\{++-\}$} and
\mbox{$\{+-+\}$} and  \mbox{$\{-++\}$} for which the general solutions
\mbox{$\{w_j\}$}  and   \mbox{$\{v_k\}$}  to  (\ref{evbae})  are quite
different. The  eigenvalues $\Lambda(v)$,   however, remain the  same,
because each  set $\{\epsilon_k\}$ describes  the same physical system
--  the $t-J$  model.  Thus,  we  may  confine  ourselves to the  case
\mbox{$\{++-\}$}     which  implies  analytical  simplifications.  The
largest eigenvalue of the quantum transfer  matrix is characterized by
$N/2$  roots for each set  \mbox{$\{w_j\}$} and \mbox{$\{v_k\}$}.  Due
to the  structure of  the Bethe ansatz   equations (\ref{evbae}) it is
natural to assume (at least for vanishing magnetic field $h$) that the
two  sets  \mbox{$\{w_j\}$}  and  \mbox{$\{v_k\}$}   for the   largest
eigenvalue are symmetric with respect to complex conjugation 
\begin{equation}
  w_j=\overline{v}_j,\quad j=1,\dots,N/2,\quad h=0.
\label{tvsym}
\end{equation}
We are led to this conjecture by the typical  situation in other Bethe
ansatz systems, e.g. the groundstate of the Heisenberg model where the
roots are symmetrically arranged in the complex plane. 

Using (\ref{tvsym})  the two different  sets of Bethe ansatz equations
(with \mbox{$h=0$}) reduce to 
\begin{equation}
  \left(\frac{v_i+{\rm i}\, u+{\rm i}}{v_i+{\rm i}\, u}\right)^{N/2}
   =-\prod_{j=1}^{N/2}\frac{v_i-\overline{v}_j+{\rm i}}{v_i-\overline{v}_j}
  \: {\rm e}^{-\beta\mu}.
  \label{tvbaed}
\end{equation}
Due to the denominator on the right-hand side no real root is allowed,
i.e.   all   roots    must  possess  non-vanishing  imaginary   parts.
Eventually, we are interested in the Trotter limit \mbox{$N\to\infty$}
in  order to calculate  thermodynamical properties.  It turns out that
for   \mbox{$N\to\infty$} the  roots  accumulate  at  the origin  with
vanishing  imaginary  (as well  as  real) parts.   This delicate point
makes it difficult to analyze  the limit \mbox{$N\to\infty$}  directly
on  the  basis of  the Bethe ansatz  equations  either analytically or
numerically.  One can   overcome  this  problem  by  introducing  well
adapted integral equations which is the topic of  the next section.  A
detailed investigation  of properties  of  the Bethe ansatz   roots is
presented in appendix~\ref{sec:roots}.

\section{Non-linear integral equations }
\label{sec:nle}

The eigenvalues $\Lambda(v)$ of  ${\cal T}^{\cal QTM}(v)$ are analytic
functions of  the spectral parameter  $v$.  We use this analyticity to
determine    the  largest  eigenvalue    with explicit  representation
(\ref{evlambda}) by   a finite  set of non-linear   integral equations
\cite{KluWe96,JuKlu96}.   This approach   also  allows for taking  the
limit $N\to\infty$  analytically.    We   found  that the    following
combinations     of    $\lambda_{\pm}(v)$  and   $\lambda_0(v)$   (see
(\ref{evlambda_part})) 
\begin{align}
  \mathfrak{b}&:= \frac{\lambda_-}{\lambda_++\lambda_0}, &
  \mathfrak{B}&:= 1+\mathfrak{b},\\
  \overline{ \mathfrak{b}}&:= \frac{\lambda_+}{\lambda_-+\lambda_0}, &
  \overline{\mathfrak{B}}&:=1+\overline{\mathfrak{b}}, \\  
  \mathfrak{c}&:=\frac{\lambda_{-}\lambda_{+}}
                      {\lambda_0(\lambda_{-}+\lambda_{+}+\lambda_0)}, &
  \mathfrak{C}&:=1+\mathfrak{c}.
  \label{defs}
\end{align}
define useful auxiliary  functions which  satisfy  a closed system  of
functional equations   (cf. Appendix B).   The function $\mathfrak{b}$
($\overline{\mathfrak{b}}$) is  an  analytic complex function  along a
finite strip in the upper  (lower) half plane, while $\mathfrak{c}$ is
an analytic complex function on the real axis.  The analyticity is due
to a cancellations of singularities among the  $\lambda$'s like in the
case  of  $\Lambda(v)$.  In  this sense,  the identities between these
auxiliary   functions  encode  the  information   on the Bethe  ansatz
equations as pointed out in the previous section. 

After   some      lengthy calculations    which   are     described in
appendix~\ref{sec:nle_app} we  obtain the following  relations for the
auxiliary functions 
\begin{equation}
  \begin{split}
  \log\mathfrak{b}(x)&=
  -2\pi\beta\,\Psi_\mathfrak{b}(x+{\rm i}\,\gamma)+\beta(\mu+h/2)\\&\quad
  -\Psi_\mathfrak{b}\ast
  \log(1+{\overline{\mathfrak{b}}})|_{{x+2{\rm i}\,\gamma}}
  -\Psi_\mathfrak{b}\ast \log(1+\mathfrak{c})|_{{x+{\rm i}\,\gamma}}, \\ 
  \log{\overline{\mathfrak{b}}}(x)&=
  -2\pi\beta\,{\Psi_{\overline{\mathfrak{b}}}}
  (x-{\rm i}\,\gamma)+\beta(\mu-h/2)\\&\quad
  -{\Psi_{\overline{\mathfrak{b}}}}\ast
  \log(1+\mathfrak{b})|_{{x-2{\rm i}\,\gamma}}
  -{\Psi_{\overline{\mathfrak{b}}}}\ast
  \log(1+\mathfrak{c})|_{{x-{\rm i}\,\gamma}}, \\ 
  \log\mathfrak{c}(x)&=
  -2\pi\beta\,\Psi_\mathfrak{c}(x)+2\beta\mu\\&\quad
  -\Psi_\mathfrak{b}\ast
  \log(1+{\overline{\mathfrak{b}}})|_{{x+{\rm i}\,\gamma}}
    -{\Psi_{\overline{\mathfrak{b}}}}\ast
    \log(1+\mathfrak{b})|_{{x-{\rm i}\,\gamma}}
    -\Psi_\mathfrak{c}\ast\log(1+\mathfrak{c})|_{{x}},
  \end{split}
\label{nle}
\end{equation}
with the driving terms (and kernels)
\begin{equation*}
    2\pi\,{\Psi_\mathfrak{b}}(x)=\frac{1}{x(x-{\rm i})},\quad
    2\pi\,{\Psi_{\overline{\mathfrak{b}}}}(x)=\frac{1}{x(x+{\rm i})},
\quad\text{and}\quad 
    2\pi\,{\Psi_\mathfrak{c}}(x)=\frac{2}{x^2+1},  
\end{equation*}
where $\ast$ denotes  the usual convolution \mbox{$f\ast g|_{{x}}=\int
  f(x-y)\,  g(y)\,{\rm d}y$}   taken at the  indicated arguments  $x$,
\mbox{$x\pm{\rm  i}\,\gamma$}  and \mbox{$x\pm 2{\rm i}\,\gamma$} with
arbitrary but fixed \mbox{$0<\gamma<1$}.  The solution of the integral
equations provides the largest eigenvalue via: 
\begin{equation}
  \log\Lambda=-\log\mathfrak{c}(0)+2\beta\mu.
\label{nle-eval}
\end{equation}
For vanishing magnetic  field ($h=0$) the functions $\mathfrak{b}(x)$,
$\overline{\mathfrak{b}}(x)$ are  related by complex conjugation which
leads  to a reduction to  two  nonlinear integral equations. For  many
applications we need only this case to study. 

The integral equations (\ref{nle}) can be solved by iteration (for the
time being \mbox{$h=0$}) 
\begin{align*}
    \mathfrak{b}^{(k+1)}&=\exp\left(
      -2\pi\beta\,\Psi_\mathfrak{b}+\beta\mu\ 
      -\Psi_\mathfrak{b}\ast\log(1+\overline{\mathfrak{b}^{(k)}})
      -\Psi_\mathfrak{b}\ast \log(1+\mathfrak{c}^{(k)})\right),\\
    \mathfrak{c}^{(k+1)}&=\exp\left(
    -2\pi\beta\,\Psi_\mathfrak{c}+2\beta\mu
    -2\Re(\Psi_\mathfrak{b}\ast
    \log(1+\overline{\mathfrak{b}^{(k)}}))
    -\Psi_\mathfrak{c}\ast\log(1+\mathfrak{c}^{(k)})
    \right).
\end{align*}
Choosing appropriate initial functions (see  next section) the  series
\mbox{$\{\mathfrak{b}^{(k)},\mathfrak{c}^{(k)}\}$}                with
\mbox{${k=0,1,2,\dots}$} converges rapidly.   In  practice only a  few
steps are necessary to reach a high-precision result.  Moreover, using
the known finite  fast Fourier transform  algorithm we can compute the
convolutions very efficiently. 

In order to calculate derivatives of the thermodynamical potential one
can avoid  numerical differentiations  by  utilizing similar  integral
equations guaranteeing the  same  numerical accuracy as for  the  free
energy. The idea is as follows. Consider the function 
\begin{equation*}
l\mathfrak{c}_{\beta}:=
\frac{\partial}{\partial\beta}\log\mathfrak{c}
\quad\text{with}\quad
\frac{\partial}{\partial\beta}\log(1+\mathfrak{c})=
\frac{1}{1+\mathfrak{c}}\frac{\partial\mathfrak{c}}{\partial\beta}=
\frac{\mathfrak{c}}{1+\mathfrak{c}}\,l\mathfrak{c}_{\beta},
\end{equation*}
we have
\begin{equation}
  l\mathfrak{c}_{\beta}=
  -\frac{\partial}{\partial\beta}(2\pi\beta\,\Psi_\mathfrak{c})
  -\Psi_\mathfrak{c}\ast\left(\frac{\mathfrak{c}}{1+\mathfrak{c}}\,
   l\mathfrak{c}_{\beta}\right)-\dots,
\label{nle-eval-derived}
\end{equation}
which is a linear  integral equation for $l\mathfrak{c}_{\beta}$.  The
equations for $l\mathfrak{b}_{\beta}$ and $l\mathfrak{c}_{\beta}$ (and
similar ones   such  as   for $l\mathfrak{b}_{\mu}$,\dots)   are  also
solvable   by iteration after  the  computation  of $\mathfrak{b}$ and
$\mathfrak{c}$.

\begin{figure}[tb]
  \begin{center}
    \leavevmode
    \includegraphics[width=0.45\textwidth]{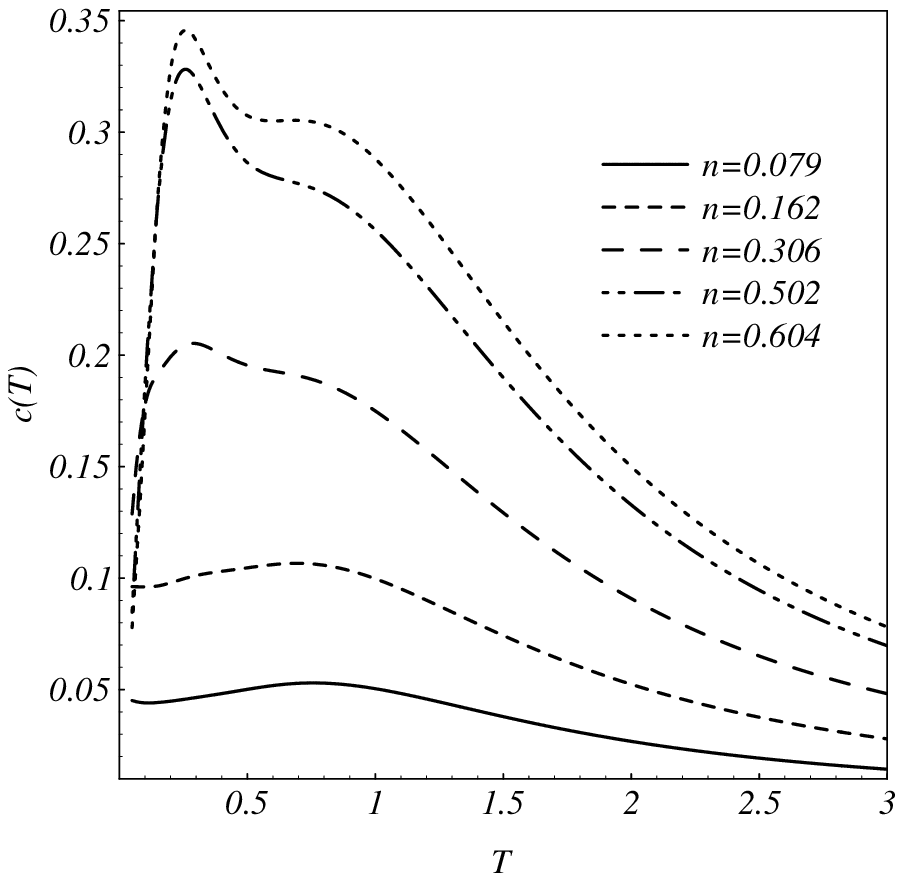}
    \includegraphics[width=0.45\textwidth]{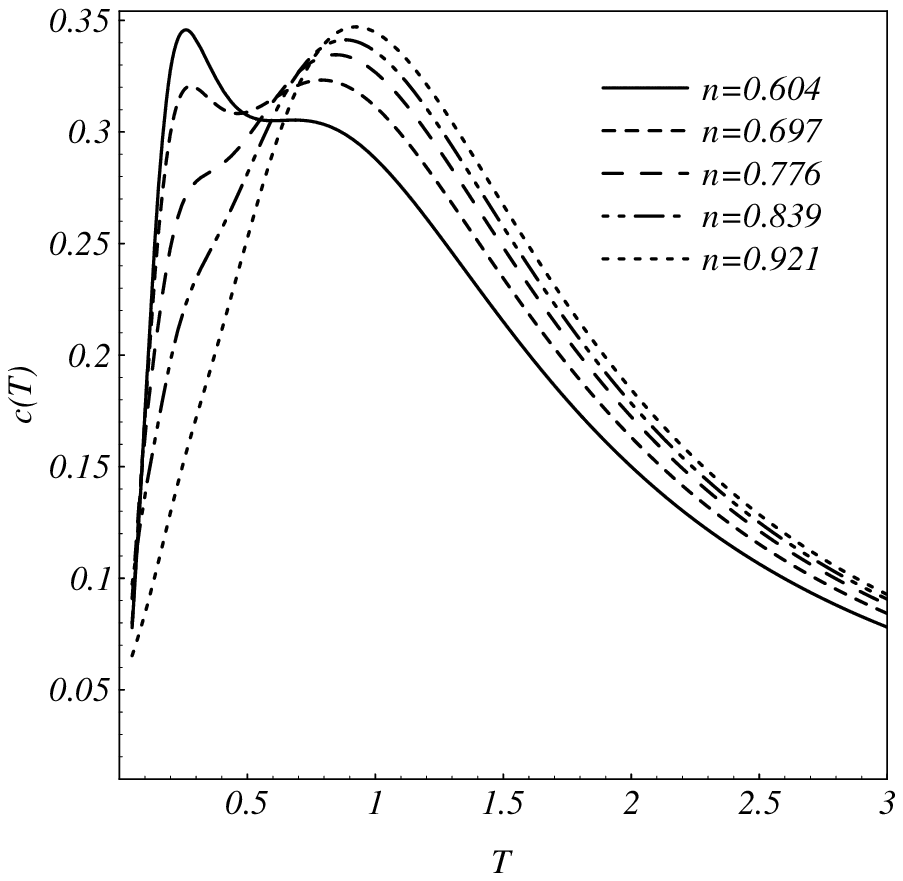} 
    \caption{Specific heat as function of $T$ (with
      \mbox{$h=0$})  for different particle  densities $n$  with 
      $n\le 0.6$ and $n\ge 0.6$.} 
    \label{fig:fig2ab}
  \end{center}
\end{figure}

Now we  consider  various thermodynamical  quantities  at intermediate
temperatures   by   solving   the    integral  equations   (\ref{nle})
numerically.  Using the   iteration   described we obtain  the   grand
canonical  potential \mbox{$f=f(T,\mu)$}  by  (\ref{nle-eval}) and its
derivative  by     (\ref{nle-eval-derived}), e.g.    the  entropy~$S$,
particle density~$n$ and the specific heat~$C$ are calculated by means
of 
\begin{equation*}
  S=-\left(\frac{\partial{f}}{\partial{T}}\right)_\mu,\quad
  n=-\left(\frac{\partial{f}}{\partial{\mu}}\right)_T,\quad
  C=T\left(\frac{\partial{S}}{\partial{T}}\right)_n.
\end{equation*}
As we are interested  in  the  thermodynamical quantities with   fixed
particle  density~$n$ we have  to   allow for a temperature  dependent
chemical potential \mbox{$\mu=\mu(T)$} which follows from 
\begin{equation*}
  \left(\frac{\partial{\mu}}{\partial{T}}\right)_n
   =-\left(\frac{\partial{n}}{\partial{T}}\right)_\mu
    \left(\frac{\partial{n}}{\partial{\mu}}\right)_T^{-1},
  \quad
  \left(\frac{\partial{S}}{\partial{T}}\right)_n
   =\left(\frac{\partial{S}}{\partial{T}}\right)_\mu
   -\left(\frac{\partial{n}}{\partial{T}}\right)_\mu^2
    \left(\frac{\partial{n}}{\partial{\mu}}\right)_T^{-1}.
\end{equation*}
From the second equation the  specific heat for fixed particle density
$n$ is  obtained   within the grandcanonical   ensemble. For numerical
results  compare  Figure~\ref{fig:fig2ab}. First  of   all, we note  a
linear temperature dependence at low $T$. According to conformal field
theory  the coefficient  is  given by  \mbox{$\pi({1/v_s}+{1/v_c})/3$}
where  $v_s$ and $v_c$  are the velocities of  the elementary spin and
charge         excitations  (see        also    (\ref{cft})         in
section~\ref{sec:low-dens}).  Our  numerical data  are consistent with
this expression.  (For a  completely analytical argument deriving  the
Luttinger liquid properties in the low-temperature limit the reader is
referred to section~\ref{sec:low-temp}). 
\begin{figure}[tb]
  \begin{center}
    \leavevmode
    \includegraphics[width=0.45\textwidth]{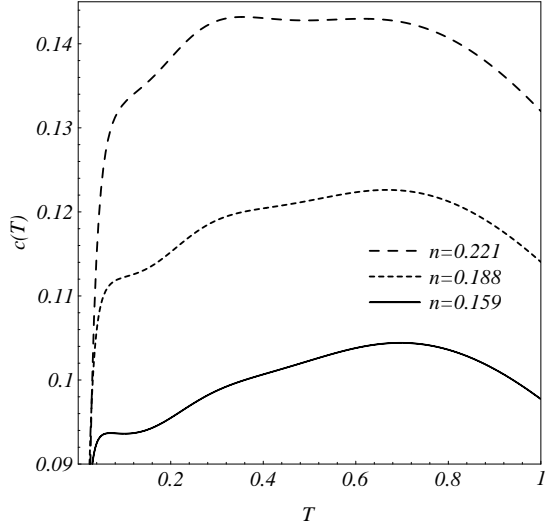} 
    \caption{Specific heat at low temperatures $T$ for intermediate 
      particle densities~$n$ and \mbox{$h=0$}.}
    \label{fig:fig3}
  \end{center}
\end{figure}
\noindent%
Furthermore, we    observe two  maxima   with changing   dominance for
increasing particle density $n$.  The nature of  this structure can be
understood  from the  elementary  excitations of  the system.  In  the
groundstate  the particles are bound  in   singlet pairs with  binding
energies varying from zero to some  density dependent value. There are
two types of excitations.  First, there  are charge excitations due to
energy-momentum  transfer onto individual   pairs.  Second, there  are
excitations due to the breaking of pairs.  The latter excitation is of
spin type at  lower excitation energies,  but changes the character at
higher (density dependent) energies to charge type as it describes the
motion of single particles. Therefore, the first and second maximum at
lower  densities (figure   \ref{fig:fig2ab}.a)  are  caused by  charge
excitations due to  pairs  and  single particles,  respectively.    At
higher  densities   (figure \ref{fig:fig2ab}.b)  the maximum  at lower
temperatures  is dominated by excitations  of pairs whereas the second
one at     higher temperatures is  caused   by  spin excitations.  For
increasing concentration the spin contribution becomes dominant as the
charge excitations freeze  out.     This is in accordance   with   the
limiting case $n=1$ leading to the spin-1/2 Heisenberg chain.  We will
come back to  this point in section~\ref{sec:high-dens}.   The missing
spin structure in the specific heat  at low and intermediate densities
is found at quite low temperatures shown in figure~\ref{fig:fig3}. 

It  is worthwhile to compare these  results  with the findings for the
Hubbard model  investigated by the  traditional thermodynamical  Bethe
ansatz \cite{UsuKawa90}.  The structure found in  the specific heat is
explained by spin and  charge  excitations which  do not  change their
character in  contrast to the $t-J$  model.   For certain  densities a
low-temperature  charge peak  was found   which  is caused by   single
particle  excitations.  A charge   peak at higher temperatures appears
because  of excitations due  to  doubly  occupied lattice  sites  with
energies of the order $U$ for large Coulomb interaction. 

To  conclude our investigation we present  numerical results for other
thermodynamical quantities.  
\begin{figure}[tb]
  \begin{center}
    \leavevmode 
    \includegraphics[width=0.45\textwidth]{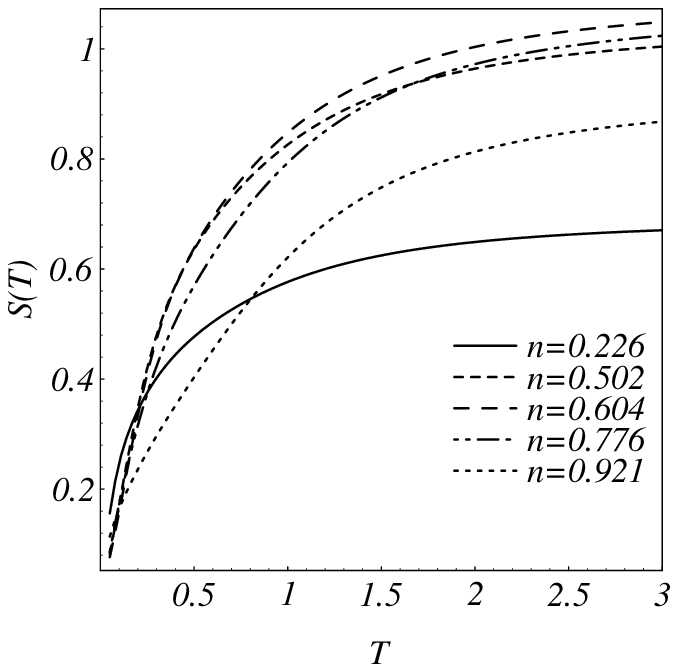}
    \includegraphics[width=0.45\textwidth]{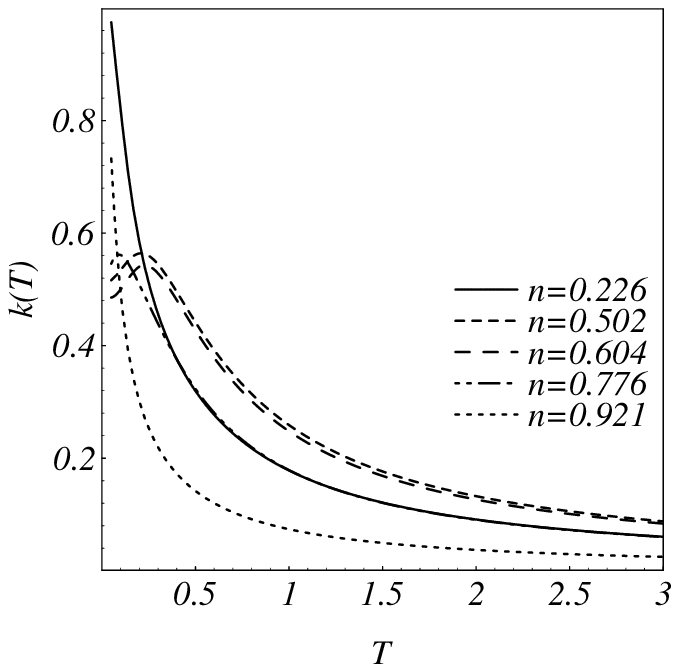}
    \caption{Entropy $S$ and compressibility $\kappa$ 
      versus $T$ for different densities~$n$.}
    \label{fig:fig4ab}
  \end{center}
\end{figure}
\noindent%
Figure~\ref{fig:fig4ab}  presents the
entropy               and             the              compressibility
\mbox{$\kappa=\partial{}n/\partial\mu$}     for     various   particle
densities.   Note the  divergent  low-temperature compressibility  for
particle densities \mbox{$n\to{0}$} and \mbox{$n\to{1}$}. 

\section{Analytical solutions of the integral equations}
\label{sec:analytical}

\subsection{Low-density regime}
\label{sec:low-dens}

In the previous sections we have applied the algebraic Bethe ansatz to
the quantum transfer matrix  and derived non-linear integral equations
(\ref{nle}) and  (\ref{nle-eval}) for the  largest eigenvalue which is
directly related to the free energy (\ref{free-energy}) of the quantum
system at  finite   temperature \mbox{$T=1/\beta$}.  This   section is
devoted  to the low-density   limit  and analytical  solutions to  the
integral equations. 

Let us consider the  low-temperature limit for small  $\mu$, i.e.  the
case  \mbox{$\beta\gg{1}$}   with  \mbox{$\mu\ll{1}$}.  Because     of
\mbox{$2\pi\beta\,\Psi_{\mathfrak{b},\mathfrak{c}}\gg{1}$} for   small
values $x$ in (\ref{nle}), we see that the functions $\mathfrak{b}$
and  $\mathfrak{c}$   are   almost  zero.  Therefore,    the essential
contribution in (\ref{nle}) is caused  by \mbox{$x\gg{1}$}. As a first
approximation we can assume that  $\mathfrak{b}$ and $\mathfrak{c}$ do
not vary much for which the equations read (\mbox{$h=0$})
\begin{align*}
    \log\mathfrak{b}&\simeq
      -\beta/x^2+\beta\mu\ 
      -\log(1+\mathfrak{b})
      -\log(1+\mathfrak{c}),\\
    \log\mathfrak{c}&\simeq
    -2\beta/x^2+2\beta\mu
    -2\log(1+\mathfrak{b})
    -\log(1+\mathfrak{c}),  
\end{align*}
where the function $\mathfrak{b}$  becomes real for $\gamma=1/2$. As a
solution we immediately have 
\begin{equation}
  \mathfrak{b}=
  \frac{{\rm e}^{-\beta/x^2+\beta\mu}}
       {1+{\rm e}^{-\beta/x^2+\beta\mu}}
  \quad\text{and}\quad
  \mathfrak{c}=
  \frac{{\rm e}^{-2\beta/x^2+2\beta\mu}}
       {1+2{\rm e}^{-\beta/x^2+\beta\mu}},
\label{bc-solution}
\end{equation}
which is inserted in (\ref{nle}, \ref{nle-eval})  
again to obtain an approximation for
the largest  eigenvalue (\ref{nle-eval}).   The analytical  expression
for  the  grand  canonical potential \mbox{${f}=-(\log\Lambda)/\beta$}
yields (after the change of variable \mbox{$x\to\sqrt{\beta}/x$}) 
\begin{align}
f&=-\int_{-\infty}^\infty\frac{{\rm d}x}{\pi\beta^{3/2}}\, 
   \log\big(1+{\rm e}^{-x^2+\beta\mu}\big),
\label{pot-int}\\
 &=-\frac{4}{3\pi}\mu^{3/2}
   -\frac{\pi}{6\sqrt{\mu}}\, T^2+{o}(\mu^{3/2},T^2/\sqrt{\mu}), 
\label{pot}
\end{align}
in agreement  with the numerical  results.  This implies for the 
particle       density  \mbox{$n=-\partial{f}/\partial\mu$},   entropy
\mbox{$S=-\partial{f}/\partial{T}$}    and        specific        heat
\mbox{$C=T\partial{S}/\partial{T}$}: 
\begin{equation}
  n=\frac{2}{\pi}\sqrt{\mu},\quad S=\frac{2}{3n}\, T,
  \quad\text{and}\quad
  C=\frac{2}{3n}\, T.
\label{entropy}
\end{equation}
These values represent thermodynamical  quantities of the $t-J$  model
(with  \mbox{$h=0$}) for small particle  densities  \mbox{$n\ll 1$} in
the     low-temperature    limit  \mbox{$T\ll{1}$}        such    that
\mbox{$(T/n)\ll{1}$}. 

The  low-temperature   behavior   of the  $t-J$     model described by
(\ref{entropy}) can be compared with independent calculations by means
of conformal field theory \cite{BloCaNi86,Aff86}.   Due to predictions
of   conformal  field  theory   we   expect  for  the  low-temperature
asymptotics 
\begin{equation}
  f=f_0-\left(\frac{\pi c_s}{6 v_s}+\frac{\pi c_c}{6 v_c}\right)T^2,
\label{cft}
\end{equation}
where $v_{s,c}$  and $c_{s,c}$ are the  velocities and central charges
(\mbox{$c_{s,c}=1$}) for the  elementary spin  and charge excitations.
For   small particle    densities   we  have   \mbox{$v_{s,c}=\pi{n}$}
\cite{BarBlaOg91}, thus (\ref{pot}) and (\ref{cft}) are consistent. 

On   the other hand,     we can  compute  the  high-temperature  limit
\mbox{$T\to\infty$}, i.e.  \mbox{$\beta\to{0}$}.  The solutions to the
integral equations (\ref{nle}) are constants. One obtains 
\begin{equation}
  \mathfrak{b}=
  \frac{{\rm e}^{\beta\mu}}{1+{\rm e}^{\beta\mu}},\quad
  \mathfrak{c}=
  \frac{{\rm e}^{2\beta\mu}}{1+2{\rm e}^{\beta\mu}}
  \quad\text{and}\quad
  f=-\beta^{-1}\log(1+2{\rm e}^{\beta\mu}),
\end{equation}
leading to 
\begin{equation}
  S=n\log\frac{2(1-n)}{n}-\log(1-n)
  \quad\text{with}\quad
  \mu=T\log\frac{n}{2(1-n)},
\end{equation}
as expected by counting the degrees of freedom  per lattice site.  The
simple high-temperature limit  for  finite $\mu$ immediately  leads to
\mbox{$S=\ln{3}$} and \mbox{$n=2/3$}.

\begin{figure}[tb]
  \begin{center}
    \leavevmode
    \includegraphics[width=0.45\textwidth]{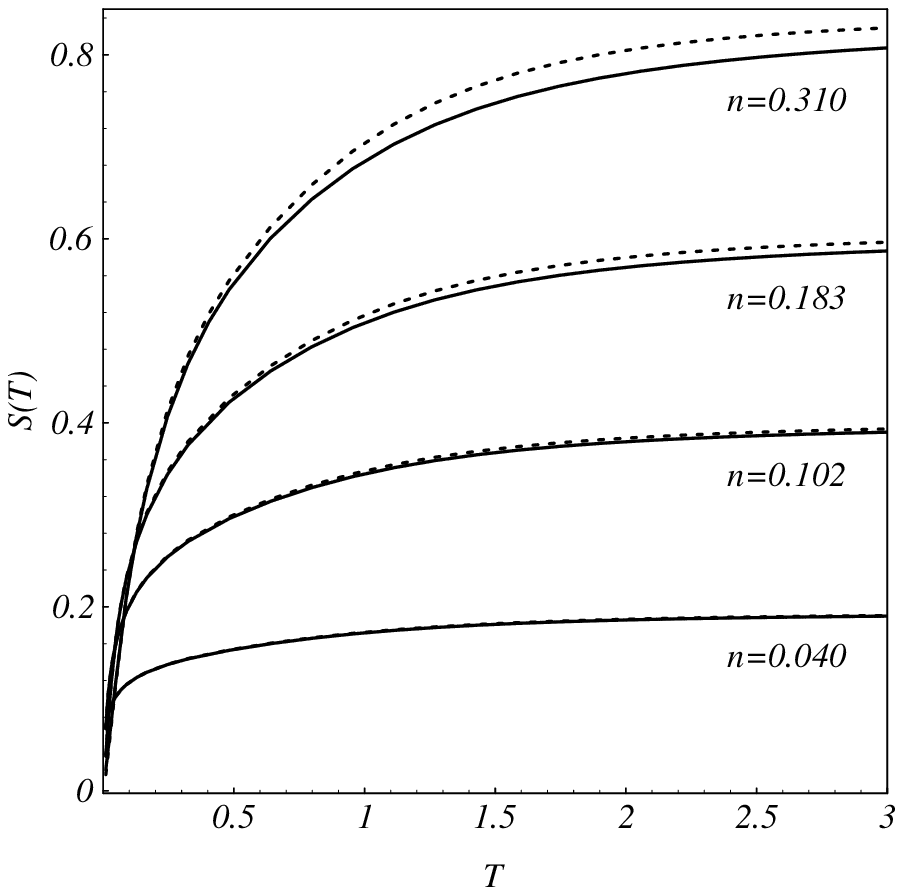}
    \includegraphics[width=0.45\textwidth]{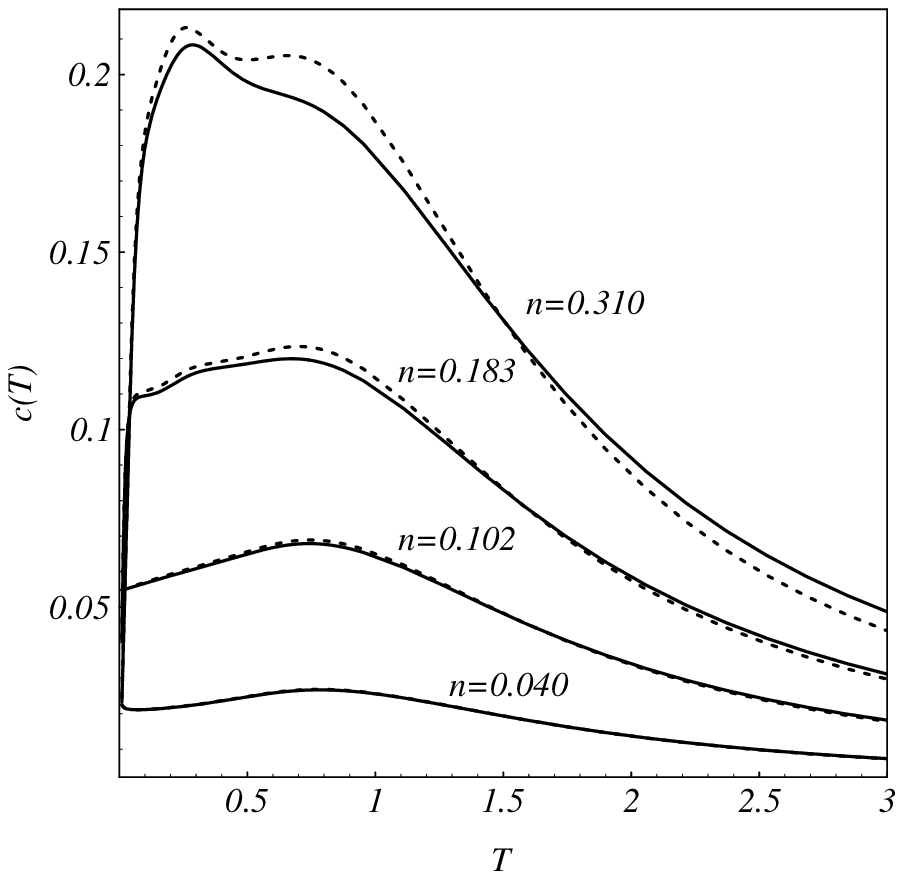} 
    \caption{Comparison of exact (numerical) calculations 
      (denoted by  the solid line)  with the  analytical approximation
      (denoted by the  dotted line) according  to (\ref{pot-finiteT}).
      Entropy $S$ (a) and specific  heat $C$ (b) as function
      of $T$ for different particle densities $n$.}
    \label{fig:fig1ab}
  \end{center}
\end{figure}
The analytical  benchmarks for \mbox{$T\to 0$} and \mbox{$T\to\infty$}
imply the consistency of   our approach.  The integral   equations for
intermediate temperatures have to be treated numerically.  However, we
have found  an excellent analytical ansatz for \mbox{$0\le{T}<\infty$}
valid in  the case  of  small particle densities  \mbox{$n\ll 1$} (cf.
figure~\ref{fig:fig1ab}).     One  can  extend  (\ref{bc-solution}) to
finite~$\beta$ (with \mbox{$\gamma=1/2$}) by 
\begin{equation}
  \mathfrak{b}_o(x)=
  \frac{{\rm e}^{-\beta/(x^2+1/4)+\beta\mu}}
       {1+{\rm e}^{-\beta/(x^2+1/4)+\beta\mu}}
  \quad\text{and}\quad
  \mathfrak{c}_o(x)=
  \frac{{\rm e}^{-2\beta/(x^2+1)+2\beta\mu}}
       {1+2{\rm e}^{-\beta/(x^2+1)+\beta\mu}},
\label{bc-solution-finiteT}
\end{equation}
yielding (\ref{nle-eval})
\begin{equation}
  f_o=-\int_{-\infty}^\infty\frac{{\rm d}x}{\pi\beta}\left(
  \frac{1}{x^2+1/4}\,
  \log\big(1+\mathfrak{b}_o(x)\big)
  +\frac{1}{x^2+1}\,
  \log\big(1+\mathfrak{c}_o(x)\big)\right).
  \label{pot-finiteT}  
\end{equation}
It  turns  out that  this    function describes  the   thermodynamical
properties of  the $t-J$ model  very well.  In figure~\ref{fig:fig1ab}
the entropy  and specific heat  (derived from (\ref{pot-finiteT})) are
depicted as  a function of $T$  and  $n$.  They  are compared with the
exact quantities calculated  numerically.   In particular,   for small
particle densities $n\ll  1$ the  deviation  from the exact  values is
negligible.   Furthermore,   the   low-temperature limit provides  the
integral   (\ref{pot-int}) with    the   corresponding     asymptotics
(\ref{pot}). 

\subsection{High-density regime}
\label{sec:high-dens}

Next, we turn to  the study of  the $t-J$ model in  the limit of large
chemical potential $\mu$ while keeping the temperature finite.  With a
glance  to  (\ref{nle}) we see   that  for large $\mu$  the  following
scaling behavior sets in 
\begin{equation*}
    \log\mathfrak{b}=O(1), \quad
\log{\overline{\mathfrak{b}}}=O(1), \quad
    \log\mathfrak{c}=\beta\mu+O(1).
\end{equation*}
In     this              limit                the        approximation
$\log\mathfrak{c}\simeq\log(1+\mathfrak{c})$ holds up to exponentially
small corrections.  Hence,  $\log\mathfrak{c}$ can be  solved from the
last equation in  (\ref{nle})  in terms of  the $\log(1+\mathfrak{b})$
and      $\log(1+{\overline{\mathfrak{b}}})$   functions. The   actual
calculation is done most conveniently in Fourier space. For details we
refer to Appendix B. The resulting relations are 
\begin{equation}
  \begin{split}
  \log\mathfrak{b}(x)&=
  -2\pi\beta\,\Phi(x+{\rm i}\,\gamma)+\beta h/2\\&\quad
  +k\ast\log(1+{{\mathfrak{b}}})|_{{x}}
  -k\ast\log(1+{\overline{\mathfrak{b}}})|_{{x+2{\rm i}\,\gamma}}, \\ 
  \log{\overline{\mathfrak{b}}}(x)&=
  +2\pi\beta\,\Phi(x-{\rm i}\,\gamma)-\beta h/2\\&\quad
  +k\ast\log(1+{\overline{\mathfrak{b}}})|_{{x}}
  -k\ast\log(1+\mathfrak{b})|_{{x-2{\rm i}\,\gamma}}, \\ 
  \log\mathfrak{c}(x)&=
  -2\pi\beta\,k(x)+\beta\mu\\&\quad
  +\Phi\ast\log(1+{{\mathfrak{b}}})|_{{x-{\rm i}\,\gamma}}
  -\Phi\ast\log(1+{\overline{\mathfrak{b}}})|_{{x+{\rm i}\,\gamma}},
  \end{split}
\label{nle-hd}
\end{equation}
with the driving term and kernel
\begin{equation*}
    {\Phi}(x)=-\frac{{\rm i}}{2}\,\frac{1}{\sinh\pi x},\quad
    k(x)=\frac{1}{2\pi}\,\int_{-\infty}^\infty
    \frac{{\rm e}^{{\rm i}\,k\, x}}{1+{\rm e}^{|k|}}\,{\rm d}k.
\end{equation*}
Finally, the corresponding eigenvalue of the quantum transfer matrix is
given by
\begin{equation*}
\log\Lambda=2\beta\log 2+\beta\mu
+\frac{{\rm i}}{2}\int_{-\infty}^\infty
 \frac{\log(1+{{\mathfrak{b}}(x+{\rm i}\,\gamma)})}
      {\sinh\pi (x+{\rm i}\,\gamma)}\,{\rm d}x
-\frac{{\rm i}}{2}\,\int_{-\infty}^\infty
 \frac{\log(1+{\overline{\mathfrak{b}}}(x-{\rm i}\,\gamma))}
{\sinh\pi (x-{\rm i}\,\gamma)}\,{\rm d}x.
\end{equation*}
According to this  result the particle density in  the limit  of large
chemical potential   is \mbox{$n=1$}  which  in fact  is   the largest
possible value.  By comparison  of (\ref{nle-hd}) with \cite{Klu93} we
find    the  nonlinear     integral   equations  of   the    isotropic
antiferromagnetic spin-1/2 Heisenberg chain.  This was expected as the
$t-J$ model in the limit of ``half-filling'' reduces to the Heisenberg
model. 

\subsection{Low-temperature asymptotics}
\label{sec:low-temp}

Let us   consider the  model  at  low  temperatures but   at arbitrary
filling.  We will employ an approximation to obtain the free energy in
the  low-temperature regime up  to  explicit $O(1/\beta)$ terms.   The
approximation yields a direct relation between the truncated nonlinear
integral  equations    and  the so-called  dressed   energy  formalism
\cite{BoIzKo86,IzKoRe89} for the groundstate  properties of quantum chains.
This correspondence itself is novel, and serves as a consistency check
of our results from the  nonlinear integral approach  on one side  and
the  dressed energy formalism and conformal  field theory on the other
side. 

Consider  the    system    in an  external    magnetic    field.  Then
$\mathfrak{b}(x)$  and   $\overline{\mathfrak{b}}(x)$ are   no  longer
complex conjugate to each other.  The latter function can be neglected
as $\overline{\mathfrak{b}}  \sim O({\rm  e}^{-\beta h})$ for positive
$h$.  We choose  $\gamma=1/2$    so   that $\mathfrak{b}(x)$  is     a
real-valued  function  on  the  real   axis.   Then the   approximated
nonlinear integral equations read 
\begin{equation}
\begin{split}
\log  \mathfrak{b}(x) &= -\beta \epsilon^0_\mathfrak{b}(x) -
   \Psi_\mathfrak{b} *\log(1+\mathfrak{c})|_{x+{\rm i}/2}, \\                
\log \mathfrak{c}(x) &= -\beta \epsilon^0_\mathfrak{c}(x) -
   \Psi_\mathfrak{b} *\log(1+\mathfrak{b})|_{x+{\rm i}/2}
                        -\Psi_\mathfrak{c} *\log(1+\mathfrak{b})|_{x},
\end{split}
\label{(J1)}
\end{equation}
where 
\begin{equation*}
  \epsilon^0_\mathfrak{b}(x)= 
  2\pi \Psi_\mathfrak{b}(x+{\rm i}/2)-(\mu+h/2)
  \quad\mbox{and}\quad
  \epsilon^0_\mathfrak{c}(x) =2\pi \Psi_\mathfrak{c}(x)-2\mu.
\end{equation*}
Numerically,    we    have  verified    that   $\mathfrak{b}(x)$   and
$\mathfrak{c}(x)$ exhibit a  crossover behavior from $\mathfrak{b}(x),
\mathfrak{c}(x) \ll 1$  to $\mathfrak{b}(x), \mathfrak{c}(x)  >1$. Due
to the driving term in (\ref{(J1)}) of  order $O(\beta)$ the crossover
becomes  very   pronounced in  the  low-temperature  limit.  We define
``Fermi   surfaces''    by   $\mathfrak{b}(\pm\Lambda_{\mathfrak{b}})=
\mathfrak{c}(\pm\Lambda_{\mathfrak{c}})=1$.  Keeping these remarks  in
mind, we   split the  contribution from the   first integral  term  in
(\ref{(J1)}) into three pieces 
\begin{align}
\begin{split}
\Psi_\mathfrak{b} *\log (1+\mathfrak{c})|_{x+{\rm i}/2}  
\quad\rightarrow\quad 
&\frac{1} {2 \pi} \int_{|x'|>\Lambda_{\mathfrak{c}}} T_{s,c}(x-x') 
\log \mathfrak{c}(x')\,\hbox{d} x'\cr
+& 
\frac{1}{2\pi} \int_{|x'|>\Lambda_{\mathfrak{c}}}  T_{s,c}(x-x')
                      \log(1+\frac{1}{\mathfrak{c}(x')})  \,\hbox{d} x' \cr
+&
\frac{1}{2\pi} \int_{-\Lambda_{\mathfrak{c}}}^{\Lambda_{\mathfrak{c}}}
T_{s,c}(x-x')
\log(1+\mathfrak{c}(x'))   \,\hbox{d} x'.
\end{split}
\label{(J2)}
\end{align}
where 
\begin{align}
\begin{split}
T_{s,c}(x)&=2\pi \Psi_\mathfrak{b}(x+{\rm i}/2)=1/(x^2+1/4),\\
T_{c,c}(x)&=2\pi \Psi_\mathfrak{c}(x) =2/(x^2+1).\\
\end{split}
\end{align}
As the slope is sufficiently steep at low temperatures, we can
approximate $\mathfrak{c}(x)(|x|<\Lambda_{\mathfrak{c}})$ and
$1/\mathfrak{c}(x) (|x|>\Lambda_{\mathfrak{c}})$ by 
$ \exp(-(\log \mathfrak{c})'(\Lambda_{\mathfrak{c}}) 
|x\pm\Lambda_{\mathfrak{c}}|)$
in the vicinity of the ``Fermi'' surfaces.
Note that $(\log \mathfrak{c})'(\Lambda_{\mathfrak{c}})$ 
is of order $\beta$. 
We can thus justify this linearization over the total integral range
for the last two integrals in (\ref{(J2)}).
Hence the last term in (\ref{(J2)}) reduces to
\begin{align}
\begin{split}
& \sim \frac{1}{2\pi}
 (T_{s,c}(x-\Lambda_{\mathfrak{c}})+
 T_{s,c}(x+\Lambda_{\mathfrak{c}}))
  \times
 \int_{0}^{\infty} 
 \log(1+{\rm e}^{-(\log \mathfrak{c})'(\Lambda_{\mathfrak{c}}) x})\,
 \hbox{d} x\cr
 &=(T_{s,c}(x-\Lambda_{\mathfrak{c}})+
 T_{s,c}(x+\Lambda_{\mathfrak{c}}))
  \frac{\pi} {24 (\log \mathfrak{c})'(\Lambda_{\mathfrak{c}})  }.
\end{split}
\label{(J3)}
\end{align}
The second term  in (\ref{(J2)}) can be treated  in the same way,  and
turns out  to be identical to  the result in (\ref{(J3)}).  Similarly,
we can   linearize the integral  equation for  $\mathfrak{c}(x)$.  The
resultant  equations are now given by  linear  integral equations over
finite integration intervals 
\begin{equation}
\begin{split}
\log\mathfrak{b}(x) = 
&-\beta\epsilon_{\mathfrak{b}}^0(x)
-(T_{s,c}(x-\Lambda_{\mathfrak{c}})+T_{s,c}(x+\Lambda_{\mathfrak{c}}))
      \frac{\pi}{12(\log \mathfrak{c})'(\Lambda_{\mathfrak{c}})}  \\
&-\frac{1}{2\pi}\int_{|x'|>\Lambda_{\mathfrak{c}}} T_{s,c}(x-x')
  \log \mathfrak{c}(x') \,\hbox{d}x' , \\
\log \mathfrak{c}(x) = 
&-\beta\epsilon_{\mathfrak{c}}^0(x)
 -(T_{s,c}(x-\Lambda_{\mathfrak{b}})+T_{s,c}(x+\Lambda_{\mathfrak{b}}))
       \frac{\pi}{12(\log \mathfrak{b})'(\Lambda_{\mathfrak{b}})}  \\
&\phantom{-\beta\epsilon_{\mathfrak{c}}^0(x)}
 -(T_{c,c}(x-\Lambda_{\mathfrak{c}})+T_{c,c}(x+\Lambda_{\mathfrak{c}}))
       \frac{\pi}{12(\log \mathfrak{c})'(\Lambda_{\mathfrak{c}})}  \\
&-\frac{1}{2\pi}\int_{|x'|>\Lambda_{\mathfrak{b}}} T_{s,c}(x-x')
  \log\mathfrak{b}(x') \,\hbox{d}x' 
 -\frac{1}{2\pi}\int_{|x'|>\Lambda_{\mathfrak{c}}} T_{c,c}(x-x')
  \log \mathfrak{c}(x') \,\hbox{d}x'.
\end{split}
\label{(J4b)}
\end{equation}
Let us introduce the short-hand notation
\begin{align}
& \left(\begin{array}{cc}     1 &       T_{s,c} \cr
               T_{s,c}&    1+T_{c,c} \end{array}\right)
 * \left(\begin{array}{c}   \log \mathfrak{b} \cr
               \log \mathfrak{c} \end{array}\right)
 =
  \left(\begin{array}{cc} \phi_{\mathfrak{b}} \cr
            \phi_{\mathfrak{c}} \end{array}\right),   
\label{(J4')} 
\end{align}
with
\begin{equation}
\begin{split}
\phi_\mathfrak{b}(x)&=-\beta\epsilon_{\mathfrak{b}}^0(x) 
-(T_{s,c}(x-\Lambda_{\mathfrak{c}})+T_{s,c}(x+\Lambda_{\mathfrak{c}}))
       \frac{\pi}{12(\log \mathfrak{c})'(\Lambda_{\mathfrak{c}})},  \\
\phi_\mathfrak{c}(x)&=-\beta\epsilon_{\mathfrak{c}}^0(x)
- (T_{s,c}(x-\Lambda_{\mathfrak{b}})+T_{s,c}(x+\Lambda_{\mathfrak{b}}))
       \frac{\pi}{12(\log \mathfrak{b})'(\Lambda_{\mathfrak{b}})}  
\label{subsidiary}\\
&\phantom{-\beta\epsilon_{\mathfrak{c}}^0(x)}\ 
- (T_{c,c}(x-\Lambda_{\mathfrak{c}})+T_{c,c}(x+\Lambda_{\mathfrak{c}}))
       \frac{\pi}{12(\log \mathfrak{c})'(\Lambda_{\mathfrak{c}})} .
\end{split}
\end{equation}
We  are  interested    in  the  evaluation   of $\log\mathfrak{c}(0)$.
Apparently, it   consists    of  two contributions,    $O(\beta)$  and
$O(1/\beta)$  which    are   contained  in    the   integral  equation
(\ref{(J4b)}).  We will  separate   these  terms by  introducing   two
further        `dressed'   functions     \mbox{$\xi_\mathfrak{b}(x)$},
\mbox{$\xi_\mathfrak{c}(x)$} satisfying 
\begin{equation}
 \left(\begin{array}{cc}     1 &       T_{s,c} \cr
               T_{s,c}&    1+T_{c,c} \end{array}\right)
 * \left(\begin{array}{c}   \xi_{\mathfrak{b}} \cr
              \xi_{\mathfrak{c}} \end{array}\right)
 =
  \left(\begin{array}{c} T_{s,c}\cr
            T_{c,c}\end{array}\right),  
\label{(J5)}
\end{equation}
where we have  adopted  the  same  abbreviation as in   (\ref{(J4')}).
After changes in  the order of integrations,  all integrands are given
by products  of  $\xi_{\mathfrak{b}}$, $\xi_{\mathfrak{c}}$  and known
functions.  As all functions appearing here are even, we arrive at the
expression: 
\begin{align}
\begin{split}
&\log \mathfrak{c}(0)  \simeq  2(\mu-1)\beta 
+ {\frac{\beta}{2\pi}} \int_{|x|>\Lambda_{\mathfrak{b}}} 
  \epsilon_{\mathfrak{b}}^0(x) \xi_\mathfrak{b}(x) \,\hbox{d}x
+ {\frac{\beta}{2\pi}} \int_{|x|>\Lambda_{\mathfrak{c}}} 
  \epsilon_{\mathfrak{c}}^0(x) \xi_\mathfrak{c}(x) \,\hbox{d}x\\
 &-\frac{\pi} {6(\log \mathfrak{c})'(\Lambda_{\mathfrak{c}})} 
  \Bigl( T_{c,c}(\Lambda_{\mathfrak{c}})
  - \frac{1}{2\pi} \int_{|x|>\Lambda_{\mathfrak{c}}}
T_{s,c}(x-\Lambda_{\mathfrak{c}})\xi_\mathfrak{b}(x) \,\hbox{d}x
  - \frac{1}{2\pi} \int_{|x|>\Lambda_{\mathfrak{c}}}
T_{c,c}(x-\Lambda_{\mathfrak{c}})\xi_\mathfrak{c}(x) \,\hbox{d}x
         \Bigr)  \\
 &-\frac {\pi}{6(\log \mathfrak{b})'(\Lambda_{\mathfrak{b}})} 
  \Bigl( T_{s,c}(\Lambda_{\mathfrak{b}})
  - \frac{1}{2\pi} \int_{|x|>\Lambda_{\mathfrak{c}}}
T_{s,c}(x-\Lambda_{\mathfrak{b}})\xi_\mathfrak{c}(x) \,\hbox{d}x
         \Bigr).
\end{split}
 \label{(J6)}
\end{align}
By   definition (\ref{(J5)}) the  contents   of the  last brackets are
nothing    but    \mbox{$\xi_{\mathfrak{c}}(\Lambda_{\mathfrak{c}})$},
\mbox{$\xi_{\mathfrak{b}}(\Lambda_{\mathfrak{b}})$}  themselves.    We
thus obtain the free energy at finite filling as 
\begin{align}
\begin{split}
f & = \frac{1} {2\pi} \int_{|x|>\Lambda_{\mathfrak{b}}} 
   \epsilon_{\mathfrak{b}}^0(x) \xi_\mathfrak{b}(x) \,\hbox{d}x
    +\frac{1} {2\pi} \int_{|x|>\Lambda_{\mathfrak{c}}} 
      \epsilon_{\mathfrak{c}}^0(x) \xi_\mathfrak{c}(x) \,\hbox{d}x\\
  & - {\frac{\pi \xi_{\mathfrak{b}}(\Lambda_{\mathfrak{b}})}
    {6\beta (\log \mathfrak{b})'(\Lambda_{\mathfrak{b}})}}
   - {\frac{\pi \xi_{\mathfrak{c}}(\Lambda_{\mathfrak{c}})}
    {6\beta (\log \mathfrak{c})'(\Lambda_{\mathfrak{c}})}}
   +o(1/{\beta^2}).
\end{split}
\label{(J.7)}
\end{align}

Now we compare our results with  those found within the dressed energy
formalism.  We quote the  result by Kawakami and Yang \cite{KawaYa91}.
Immediately  seen  from  their  equation  (2.26),  the dressed  energy
functions for  spinon ($\epsilon_s$) and holon  ($\epsilon_c$) satisfy
the  same integral equations (\ref{(J4')})  except that the right-hand
side       should                   be                replaced      by
\mbox{$(-\epsilon_{\mathfrak{b}}^0,-\epsilon_{\mathfrak{c}}^0)$}.
Therefore, we have the connecting relations 
\begin{equation}
\log \mathfrak{b}(x) = \beta \epsilon_s(x) + O(1/{\beta}),  \qquad
\log \mathfrak{c}(x) = \beta \epsilon_c(x) + O(1/{\beta}),   
\label{(J8)}
\end{equation}
within the $O(1/\beta)$ approximation.  Now  that we want to  evaluate
$\log\mathfrak{b}(x)$, $\log\mathfrak{c}(x)$  and  $\beta f$ including
all     $O({1}/{\beta})$      corrections     we         can   replace
\mbox{$(\log\mathfrak{b})'(\Lambda_{\mathfrak{b}})$}               and
\mbox{$(\log\mathfrak{c})'(\Lambda_{\mathfrak{c}})$}      in       the
denominators               of              (\ref{(J.7)})            by
\mbox{$\beta\epsilon'_s(\Lambda_{\mathfrak{b}})$}                  and
\mbox{$\beta\epsilon'_c(\Lambda_{\mathfrak{c}})$},       respectively.
Moreover,  we find that our $\xi$   functions are proportional to bulk
density          functions         of       spinons     and    holons:
\mbox{$\xi_{\mathfrak{b},\mathfrak{c}}(x)= 2\pi   \rho_{s,c}(x)$}.  To
this end compare equation (\ref{(J5)}) with (2.20) in \cite{KawaYa91}.
Hence the integrands of the first two  terms in equation (\ref{(J.7)})
are products  of bare energy  functions times  bulk density functions.
Therefore, they give ``bulk'' (or  zero temperature) contributions, as
expected.   The  remaining terms,  which represent  finite temperature
contributions, can be neatly written  down by adopting the expressions
for the sound velocities of the  elementary excitations in the dressed
energy     approach.     Namely, let      the   sound  velocities   be
$v_{\alpha}=\epsilon_{\alpha}'/2\pi\rho_{\alpha}|_{\Lambda_\alpha}$
(with $\alpha=s,c$) then  the $O(1/\beta^2)$ terms  in the free energy
read 
\begin{equation}
 -  \frac{\pi}{6 \beta^2} \Bigl( \frac{1}{v_s}+\frac{1}{v_c} \Bigr).
\label{(J9)}
\end{equation}
We thus observe various consistency  relations.  The free energy obeys
the prediction by conformal invariance and Luttinger liquid theory for
arbitrary filling.  The sound velocities  coincide with those obtained
from the dressed energy calculations.  We  remark that there should be
some subtleties in treating the model for vanishing external fields in
the same approximation  scheme, while the original non-linear integral
equations are  valid for  all cases.   Even   for the linear  integral
equations obtained in this  section analytic solutions cannot be found
for general filling.  Therefore,  one   has to resort   to   numerical
treatments. Here    however,  we restrict   ourselves to   the already
presented  numerical  results   for  the original nonlinear   integral
equations which are not limited to the low-temperature regime.

\section{Discussion}
\label{sec:conclusion}

We have derived eigenvalue  equations for the quantum transfer  matrix
of the $t-J$ model.  This approach permits  the {\it exact} analytical
as  well  as   numerical calculation  of  thermodynamical  quantities.
Instead of solving an infinite   set of integral   equations -- as  is
necessary  in the traditional  thermodynamical Bethe ansatz -- we have
to solve integral equations for only {\it  two} functions (in the case
of vanishing magnetic field).  We have considered analytically certain
low- and  high-temperature limits  verifying the  values  predicted by
conformal   field  theory.    Moreover,    the case   of  intermediate
temperatures was treated numerically.  As shown, the specific heat and
compressibility  display  an  interesting  behavior  in  dependence on
particle density   and   temperature.   Also, the  reduction    to the
Heisenberg model in  the limit of `half-filling' has  been proved in a
straightforward way.  A direct   relation has been established to  the
dressed  energy  formulation  concerning the  groundstate  properties.
This has   been easily  achieved by   a  linearization scheme   in the
vicinity of the ``Fermi surface''.

Let  us emphasize  again  the advantage of the  novel  method.  In the
traditional  approach one has   to deal  with infinitely many  coupled
integral equations.  Except  for special cases,  the set  of equations
defies  further  analysis because of  its complexity.   Indeed, it was
almost 20 years after  the  derivation of equations \cite{Tak72}  that
the  numerical  calculation has  been  done  for several thermodynamic
quantities of the Hubbard model \cite{UsuKawa90}.  Still, the explicit
calculation allowed for only 2  (!)  bound charge rapidities and 15-30
spin  rapidities, while the original  equations contain $\infty \times
\infty$  rapidities.   Although it  was  claimed  that such truncation
works  for very low  temperatures,  it might  be rather inaccurate  at
finite temperatures unless the  Coulomb  interaction takes very  large
values. 

Our  treatment,   on the  other  hand,  deals  with only  two integral
equations which clearly is advantageous in practical calculations.  In
view of this,  the  presented work  might  be considered as  the first
approach  yielding explicit and concrete  results for a lattice system
with interacting spin-1/2 fermions   {\it  at all temperatures}.    We
actually   performed numerical  computations over   a   wide range  of
densities   and  temperatures.   The  results   are of  high  accuracy
($10^{-6}$) in the entire range of parameters. 

The application of the present approach is not limited to the study of
1D quantum chain  problems but  also  to those of deformed   conformal
field theories.   At the moment   the main tool  in  this field is the
so-called TBA  method which originates   from the  traditional  string
approach. As remarked by several authors \cite{DeVe92,BaLuZa96}, 
the novel approach is
an   efficient alternative for  scalar  models (the sine-Gordon model,
spin $1/2$ Heisenberg chain).  The  necessity  of a generalization  of
this method to  the  multicomponent case (Luttinger  liquid)  has been
stated as an important problem in \cite{ZamoPoly94}.  This program has
been carried  out in the present  paper.  It was shown  explicitly for
the  $t-J$  model.   The success   of  our  approach is certainly  not
accidental, i.e. the applicability  is not model dependent, but rather
universal  as shown   by the general    embedding of  the QTM  into  a
commuting family of matrices.  We expect  to report in the near future
further research on other models, e.g.  on a generalized Hubbard model
\cite{JuKluSu96}, 
as well as on additional thermodynamical quantities such
as correlation lengths.

\section*{Acknowledgments}

The  authors  acknowledge  financial   support by  the   {\it Deutsche
  Forschungsgemeinschaft}  under   grant  No.~Kl~645/3-1   and  by the
research  program    of   the   {\it    Sonderforschungsbereich   341,
  K\"{o}ln-Aachen-J\"{u}lich}.     The authors like to thank T.~Wehner
and J.~Zittartz for helpful discussions.

\begin{appendix}

\section{Properties of Bethe ansatz roots}
\label{sec:roots}

We introduce the function $\mathfrak{a}(x)$
\begin{equation}
  \mathfrak{a}(v)=\left(\frac{v+{\rm i}\, u+{\rm i}}
                             {v+{\rm i}\, u}\right)^{N/2}
  \prod_{j=1}^{N/2}\frac{v-\overline{v}_j}{v-\overline{v}_j+{\rm i}}
  \: {\rm e}^{\beta\mu}.
  \label{func_a_def}
\end{equation}
According to (\ref{tvbaed}) a root $v_k$ yields
\begin{equation}
  \mathfrak{a}(v_k)=-1.
  \label{rootdef}
\end{equation}
It is  useful   to understand  the   principal behavior of  the  roots
determining the largest eigenvalue.  This can be done analytically for
the case of finite \mbox{$N$}  for \mbox{$\mu=h=0$} and large $\beta$.
Within    this        limit   the      integral      equation     (see
appendix~\ref{sec:nle_app},       equations    (\ref{nle_app})     and
(\ref{kernel-bb})) can be used to obtain 
\begin{equation*}
    \log\mathfrak{a}(v_k)=(2k+1)\pi {\rm i}
    \approx\overline{\Phi^{(N)}_\mathfrak{b}(\overline{v}_k)},
\end{equation*}
providing 
\begin{equation}
  \label{RootArrangeFinite}
  v_k\simeq -\frac{{\rm i}}{2}+
     {\rm i}\,\sqrt{1/4+u^2+{\rm i}\, u\cot((2k+1)\pi/N)}, \quad
  k=0,\pm 1,\pm 2,\dots
\end{equation}
which   will be used as initial   values for $\mu$  and $\beta$ finite
(\mbox{$u=-\beta/N$}).     Numerically,  these values   are  shown  to
approximate the actual roots rather well even  for finite $\beta$.  In
the limit  \mbox{$N\to\infty$}  the equation (\ref{RootArrangeFinite})
reads 
\begin{equation}
  v_k\simeq -\frac{{\rm i}}{2}+{\rm i}\,
\sqrt{1/4-{\rm i}\,\beta/((2k+1)\pi)}.
  \label{RootArrangeInfinite}  
\end{equation}
Here one can see easily the accumulation of roots for
\mbox{$|k|{\gg}1$}. 

\begin{figure}[tbp]
  \begin{center}
    \leavevmode 
    \includegraphics[width=0.45\textwidth]{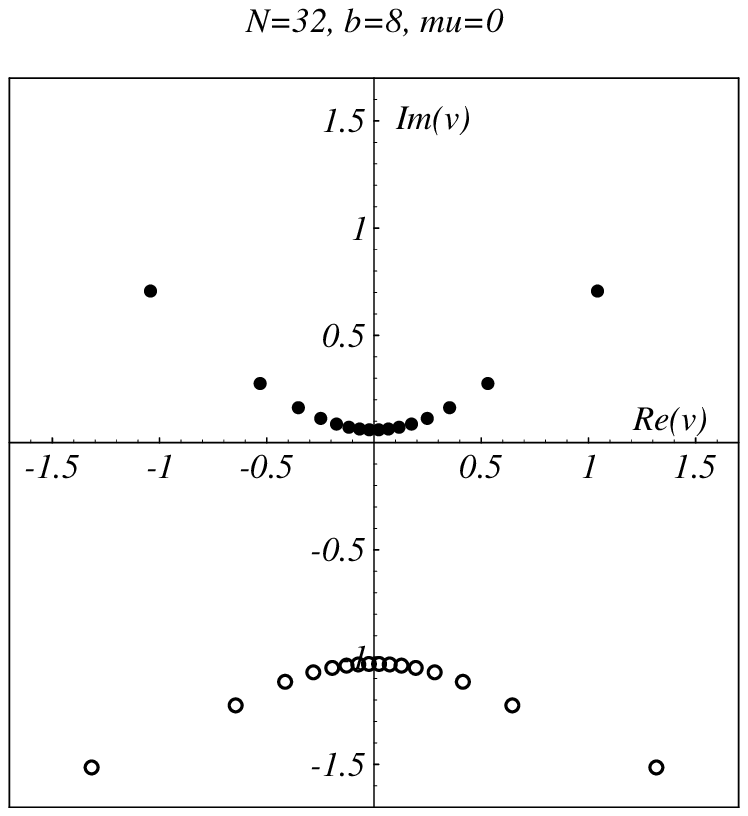}
    \includegraphics[width=0.45\textwidth]{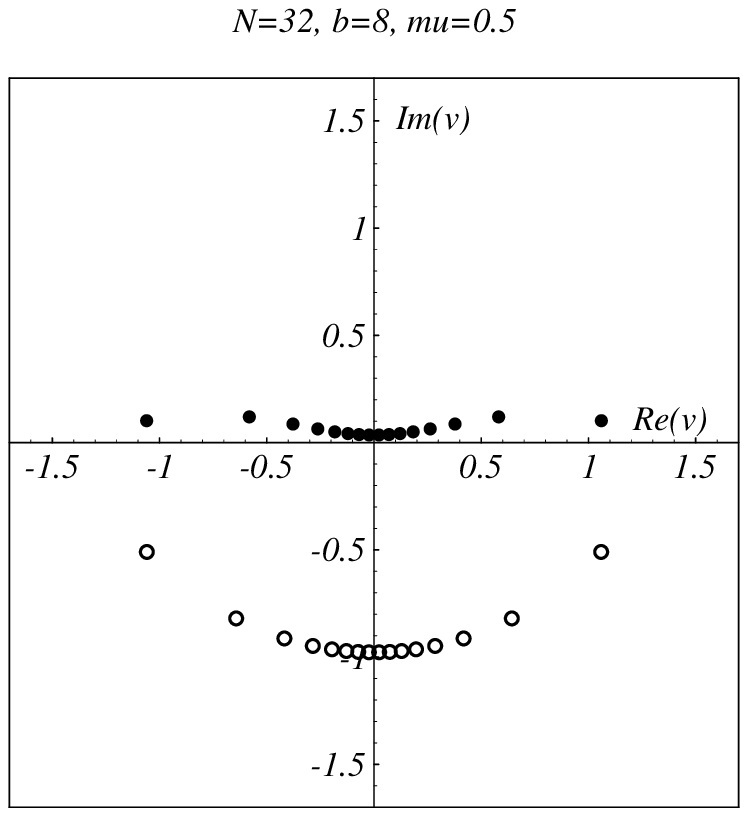}
    \caption{Distribution of roots $v_k$ (full circles) and holes
      $v_k^h$  (open circles)    in  the complex  plane    for $N=32$,
      $\beta=8$, $h=0$ and for (a) $\mu=0$ and (b) $\mu=0.5$.}
    \label{fig:fig5ab}
  \end{center}
\end{figure}
In figure \ref{fig:fig5ab} the   distribution of roots in  the complex
plane is depicted for  two different chemical  potentials  $\mu$.  The
exact values are computed numerically  by the Newton method using  the
initial  values   (\ref{RootArrangeFinite}).   Also   shown  are   the
so-called holes $v^h_k$ satisfying 
\begin{equation}
  \mathfrak{a}(v^h_k)=-1,
   \label{holedef}
\end{equation}
i.e. (\ref{rootdef}), but not coinciding with any  roots. As seen from
figure   \ref{fig:fig5ab}  the  imaginary parts   of  the roots become
smaller    for increasing $\mu$.    Simultaneously,  the largest holes
converge to the real axis.

\section{Derivation of the integral equations}
\label{sec:nle_app}

In  this appendix we describe  the detailed derivation of the integral
equations introduced in section~\ref{sec:nle}. We define the following
auxiliary functions
\begin{align}
  \mathfrak{a}&:=\frac{\lambda_0}
                      {\lambda_{+}},&
  \mathfrak{A}&:=1+\mathfrak{a}
                =\frac{\lambda_{+}+\lambda_0}
                      {\lambda_{+}},\label{a-def}\\
  \mathfrak{b}&:=\frac{\lambda_{-}}
                      {\lambda_{+}+\lambda_0},&
  \mathfrak{B}&:=1+\mathfrak{b}
                =\frac{\lambda_{-}+\lambda_{+}+\lambda_0}
                      {\lambda_{+}+\lambda_0},\label{b-def}\\
  \mathfrak{c}&:=\frac{\lambda_{-}\lambda_{+}}
                      {\lambda_0(\lambda_{-}+\lambda_{+}+\lambda_0)},&
  \mathfrak{C}&:=1+\mathfrak{c}
                =\frac{(\lambda_{-}+\lambda_0)(\lambda_{+}+\lambda_0)}
                      {\lambda_0(\lambda_{-}+\lambda_{+}+\lambda_0)},
                 \label{c-def}
\end{align}
where $\lambda_i$ are  given in (\ref{evlambda_part}).   Note that the
rescaled eigenvalue (corresponding  to   zero groundstate energy   for
vanishing chemical potential and magnetic field) 
$$\log\Lambda\rightarrow\log\Lambda-N\,\log(1-u),$$  
(with \mbox{$u=-\beta/N$})  of  the  QTM  is given by 
\begin{equation}
  \log\Lambda=N\,\log(1+u)-N\,\log(1-u)-\log(\mathfrak{c}(0))+2\beta\mu.
\end{equation}
For further convenience we write
\begin{equation}
  \Phi_\pm(v)=(v\pm {\rm i}\, u)^{N/2},\quad
  q_{-}(v)=\prod_i(v-w_i),\quad q_{+}(v)=\prod_i(v-v_i),
\end{equation}
which provide the following relations
\begin{align}
  \lambda_{-}&=\frac{q_{-}(v+{\rm i})}{q_{-}(v)}\,
  \Phi_+(v)\,\Phi_-(v-{\rm i}),\\
  \lambda_{+}&=\frac{q_{+}(v-{\rm i})}{q_{+}(v)}\,
  \Phi_-(v)\,\Phi_+(v+{\rm i}),\\
  \lambda_0&=\frac{q_{-}(v+{\rm i})}{q_{-}(v)}\, 
             \frac{q_{+}(v-{\rm i})}{q_{+}(v)}\,\Phi_+(v)\,\Phi_-(v).
\end{align}
Introducing   $\overline{\mathfrak{a}}$,    $\overline{\mathfrak{b}}$,
$\overline{\mathfrak{A}}$,   $\overline{\mathfrak{B}}$     as       in
(\ref{a-def}-\ref{c-def}) with all subscripts $+$ and $-$ interchanged
we find 
\begin{align}
  &\lambda_{-}+\lambda_0
  =\lambda_0\,\mathfrak{C}\,\mathfrak{B}
  =\lambda_{+}/\overline{\mathfrak{b}},\\
  &\lambda_{+}+\lambda_0
  =\lambda_0\,\mathfrak{C}\,\overline{\mathfrak{B}}
  =\lambda_{-}/\mathfrak{b},\label{lam_aux}\\
  &(\lambda_{-}+\lambda_0)(\lambda_{+}+\lambda_0)
  =\lambda_{-}\lambda_{+}\,\mathfrak{C}/\mathfrak{c}.
\label{lam_comb}
\end{align}
Defining the auxiliary function
\begin{equation}
  \begin{split}
\mathfrak{D}
&:=\frac{1}{q_{+}(v)}\,\big(q_{-}(v)\,
\Phi_+(v+{\rm i})+q_{-}(v+{\rm i})\,\Phi_+(v)\big),\\
&=(\lambda_{+}(v)+\lambda_0(v))
\frac{q_{-}(v)}{q_{+}(v-{\rm i})\,\Phi_-(v)}
 =\frac{q_{-}(v)}{q_{+}(v)}\,\Phi_+(v+{\rm i})\,\mathfrak{A}(v),
 \end{split}
\label{D-def}
\end{equation}
we can show that $\mathfrak{D}$ is  analytic and non-zero in the upper
complex plane ${\mathbb  C}^+$ including the real  axis.  The zeros of
the  denominator $q_{+}(v)$  cancel with  the  zeros of $\mathfrak{A}$
because these values are identical to the roots $v_i$.  Therefore, the
only  zeros   of $\mathfrak{D}$    are   caused  by  the   holes    in
${\mathbb{C}}^-$                (cf.          figure~\ref{fig:fig5ab},
appendix~\ref{sec:roots}).  It is  useful to rewrite $\mathfrak{D}$ by
means of (\ref{lam_aux}) 
\begin{align}
  \mathfrak{D}&=\frac{q_{-}(v+{\rm i})}{q_{+}(v)}\,\Phi_+(v)\,
                \mathfrak{C}(v)\,\overline{\mathfrak{B}}(v),
    \label{D-lower}\\
              &=\frac{q_{-}(v+{\rm i})}{q_{+}(v-{\rm i})}\,
  \Phi_-(v-{\rm i})\,
  \frac{\Phi_+(v)}{\Phi_-(v)}\,\frac{1}{\mathfrak{b}(v)}
    \label{D-upper},
\end{align}
which yields (\ref{lam_comb})
\begin{equation}
  \mathfrak{D}\,\overline{\mathfrak{D}}=
  \Phi_-(v-{\rm i})\,\Phi_+(v+{\rm i})\,\mathfrak{C}/\mathfrak{c}.
    \label{D-combine}
\end{equation}
Now  we use the Fourier   transform $\widehat{f}^{(\pm)}$ 
of the logarithmic derivative of $f(v)$ 
\begin{equation}
  \widehat{f}^{(\pm)}=\int_{{\cal L}^\pm}\frac{{\rm d}v}{2\pi}
                   \Big[\log f(v)\Big]^\prime {\rm e}^{-{\rm i}kv},
\end{equation}
where the integration contour ${\cal  L}^\pm$ is a straight line  near
the real  axis taken in the  upper and lower  half plane such that the
contour   is parameterized  by   \mbox{$v=x{\pm}{\rm i}\gamma$}   with
\mbox{$v\in{\cal{L}}^\pm$},       \mbox{$x\in{\mathbb{R}}$}        and
\mbox{$\gamma>|u|$}.  $\widehat{\mathfrak{D}}^{(-)}$ is calculated for
expression  (\ref{D-lower}) and   $\widehat{\mathfrak{D}}^{(+)}$   for
expression (\ref{D-upper}).  The main idea  of this procedure consists
in the implicit determination of the unknown roots,  i.e. the zeros of
$\mathfrak{A}$, by employing  functional equations in  general and the
analyticity  of  $\mathfrak{D}$  in   ${\mathbb{C}}^+$  in particular.
Provided   a     closed   set   of  equations     for  $\mathfrak{b}$,
$\overline{\mathfrak{b}}$,  and   $\mathfrak{c}$    in  terms       of
$\mathfrak{B}$, $\overline{\mathfrak{B}}$, and $\mathfrak{C}$ is found
the  eigenvalue $\Lambda$ of   the QTM can  be calculated.   Employing
(\ref{D-def}), (\ref{D-lower}) and (\ref{D-upper}) we have 
\begin{equation}
  \begin{split}
    \widehat{\mathfrak{D}}^{(-)}_{k>0}
    &={\rm e}^{-k}\,\widehat{q}_{-}+\widehat{\mathfrak{C}}
      +\widehat{\overline{\mathfrak{B}}},\\
    \widehat{\mathfrak{D}}^{(+)}_{k>0}
    &={\rm e}^{-k}\,
      \widehat{q}_{-}+\widehat{\Phi}_{+}-\widehat{\Phi}_{-}
      -\widehat{\mathfrak{b}},
  \end{split}
\qquad
  \begin{split}
     \widehat{\mathfrak{D}}^{(-)}_{k<0}
     &=\widehat{\Phi}_{+}-\widehat{q}_{+}+\widehat{\mathfrak{C}}
      +\widehat{\overline{\mathfrak{B}}}=0,\\
     \widehat{\mathfrak{D}}^{(+)}_{k<0}
     &={\rm e}^{k}\,\widehat{\Phi}_{-}-{\rm e}^{k}\,\widehat{q}_{+}
      -\widehat{\mathfrak{b}}=0.
  \end{split}
\label{D-23}
\end{equation}
where we used the  vanishing of the Fourier transform $\widehat{F}(k)$
for  \mbox{$k<0$} (\mbox{$k>0$})        for   $F(v)$  analytic      in
${\mathbb{C}}^+$   (${\mathbb{C}}^-$).   Similar   relations can    be
obtained for $\overline{\mathfrak{D}}$ reading 
\begin{equation}
  \begin{split}
     \widehat{\overline{\mathfrak{D}}}^{(-)}_{k>0}
     &={\rm e}^{-k}\,\widehat{\Phi}_{+}-{\rm e}^{-k}\,\widehat{q}_{-}
      -\widehat{\overline{\mathfrak{b}}}=0,\\
     \widehat{\overline{\mathfrak{D}}}^{(+)}_{k>0}
     &=\widehat{\Phi}_{-}-\widehat{q}_{-}+\widehat{\mathfrak{C}}
      +\widehat{\mathfrak{B}}=0,
  \end{split}
  \qquad
  \begin{split}
    \widehat{\overline{\mathfrak{D}}}^{(-)}_{k<0}
    &={\rm e}^{k}\,\widehat{q}_{+}+
    \widehat{\Phi}_{-}-\widehat{\Phi}_{+}
    -\widehat{\overline{\mathfrak{b}}},\\
    \widehat{\overline{\mathfrak{D}}}^{(+)}_{k<0}
    &={\rm e}^{k}\,\widehat{q}_{+}+\widehat{\mathfrak{C}}
    +\widehat{\mathfrak{B}}.    
  \end{split}
\label{D-13}
\end{equation}
According to (\ref{D-combine}) we have
\begin{equation}
    \widehat{\mathfrak{D}}^{(\pm)}_{k>0}
    +\widehat{\overline{\mathfrak{D}}}^{(\pm)}_{k>0}
    ={\rm e}^{-k}\,\widehat{\Phi}_{+}+\widehat{\mathfrak{C}}
      -\widehat{\mathfrak{c}},
    \qquad
    \widehat{\mathfrak{D}}^{(\pm)}_{k<0}
    +\widehat{\overline{\mathfrak{D}}}^{(\pm)}_{k<0}
    ={\rm e}^{k}\,\widehat{\Phi}_{-}+\widehat{\mathfrak{C}}
      -\widehat{\mathfrak{c}}.
\label{D-00}
\end{equation}
Consider        the        case     $k>0$.     With      respect    to
$\widehat{\mathfrak{D}}^{(-)}$     the      equations    (\ref{D-23}),
(\ref{D-13}) and (\ref{D-00}) provide 
\begin{equation}
  {\rm e}^{-k}\,\widehat{q}_{-}+\widehat{\overline{\mathfrak{B}}}
  ={\rm e}^{-k}\,\widehat{\Phi}_{+}-\widehat{\mathfrak{c}},\qquad
  \widehat{\mathfrak{c}}
  =\widehat{\overline{\mathfrak{b}}}
  -\widehat{\overline{\mathfrak{B}}}.
\end{equation}
After     inserting  the     second    relation    (\ref{D-13})    for
$\widehat{\mathfrak{D}}^{(+)}$ it follows 
\begin{align*}
  \widehat{\overline{\mathfrak{b}}}
  &={\rm e}^{-k}\,(\widehat{\Phi}_{+}-\widehat{\Phi}_{-})
   -{\rm e}^{-k}\,(\widehat{\mathfrak{B}}
          +\widehat{\mathfrak{C}}),\\
  \widehat{\mathfrak{c}}
  &={\rm e}^{-k}\,(\widehat{\Phi}_{+}-\widehat{\Phi}_{-})
   -{\rm e}^{-k}\,(\widehat{\mathfrak{B}}
          +\widehat{\mathfrak{C}})-\widehat{\overline{\mathfrak{B}}}.
\end{align*}
By similar steps we obtain the full system of equations
\begin{equation}
\begin{split}
  \widehat{\mathfrak{b}}&=
  \begin{cases}
    (\widehat{\Phi}_{+}-\widehat{\Phi}_{-})
    -(\widehat{\overline{\mathfrak{B}}} 
     +\widehat{\mathfrak{C}}) &:\quad k>0,\\
     {\rm e}^{k}\,(\widehat{\Phi}_{-}-\widehat{\Phi}_{+})
    -{\rm e}^{k}\,(\widehat{\overline{\mathfrak{B}}} 
          +\widehat{\mathfrak{C}}) &:\quad k<0, 
  \end{cases}\\
  \widehat{\overline{\mathfrak{b}}}&=
  \begin{cases}
     {\rm e}^{-k}\,(\widehat{\Phi}_{+}-\widehat{\Phi}_{-})
    -{\rm e}^{-k}\,(\widehat{{\mathfrak{B}}} 
          +\widehat{\mathfrak{C}}) &:\quad k>0,\\
     (\widehat{\Phi}_{-}-\widehat{\Phi}_{+})
    -(\widehat{{\mathfrak{B}}} 
     +\widehat{\mathfrak{C}}) &:\quad k<0, 
  \end{cases}\\
  \widehat{\mathfrak{c}}&=
  \begin{cases}
     {\rm e}^{-k}\,(\widehat{\Phi}_{+}-\widehat{\Phi}_{-})
    -{\rm e}^{-k}\,(\widehat{\mathfrak{B}}
           +\widehat{\mathfrak{C}})
    -\widehat{\overline{\mathfrak{B}}}&:\quad k>0,\\
     {\rm e}^{k}\,(\widehat{\Phi}_{-}-\widehat{\Phi}_{+})
    -{\rm e}^{k}\,(\widehat{\overline{\mathfrak{B}}}
          +\widehat{\mathfrak{C}})
    -\widehat{\mathfrak{B}}&:\quad k<0.
  \end{cases}
\end{split}
\label{Fexp}
\end{equation}
Furthermore, we obtain: 
\mbox{$(\widehat{\Phi}_{+}-\widehat{\Phi}_{-})=
{\rm i}N\,{\rm sign}(k)\,\sinh(ku)$}.
Applying the inverse Fourier transform we are  led to a system of
non-linear integral equations (denoted by the convolution $\ast$)
\begin{equation}
  \begin{split}
    \log\mathfrak{b}&=
    \Phi^{(N)}_\mathfrak{b}
    -{\Psi_\mathfrak{b}}\ast\log\overline{\mathfrak{B}}
    -{\Psi_\mathfrak{b}}\ast\log\mathfrak{C}+\beta(\mu+h/2),\\
    \log{\overline{\mathfrak{b}}}&=
    \Phi^{(N)}_{\overline{\mathfrak{b}}}
    -{\Psi_{\overline{\mathfrak{b}}}}\ast\log{\mathfrak{B}}
    -{\Psi_{\overline{\mathfrak{b}}}}\ast\log\mathfrak{C}+
     \beta(\mu-h/2),\\
    \log\mathfrak{c}&=
    \Phi^{(N)}_\mathfrak{c}
    -{\Psi_\mathfrak{b}}\ast\log\overline{\mathfrak{B}}
    -{\Psi_{\overline{\mathfrak{b}}}}\ast\log\mathfrak{B} 
    -{\Psi_\mathfrak{c}}\ast\log\mathfrak{C}+2\beta\mu,\\
  \end{split}
\label{nle_app}
\end{equation}
with
\begin{align}
  \Phi^{(N)}_\mathfrak{b}&=
  \phantom{-}{\rm i}\,N
  \left(\arctan\frac{v-{\rm i}}{u}-\arctan\frac{v}{u}\right),&
  \Psi_\mathfrak{b}&=
  \frac{1}{2\pi v(v-{\rm i})},\cr
  {\Phi^{(N)}_{\overline{\mathfrak{b}}}}&=
  -{\rm i}\,N\left(\arctan\frac{v+{\rm i}}{u}-
   \arctan\frac{v}{u}\right),&
  {\Psi_{\overline{\mathfrak{b}}}}&=
  \frac{1}{2\pi v(v+{\rm i})},\label{kernel-bb}\cr
  \Phi^{(N)}_\mathfrak{c}&=
  \phantom{-}{\rm i}\,N
  \left(\arctan\frac{v-{\rm i}}{u}-
   \arctan\frac{v+{\rm i}}{u}\right),&
  \Psi_\mathfrak{c}&=
  \frac{2}{2\pi (v^2+1)}.
\end{align}
The  additional  terms    $\beta(\mu\pm  h/2)$  and  $2\beta\mu$    in
(\ref{nle_app})    are   integration constants    which    follow from
considering the limit $v\to\infty$ 
\begin{equation}
\mathfrak{b}\to\frac{{\rm e}^{\beta(\mu+h/2)}}
                    {{\rm e}^{\beta(\mu-h/2)}+1},\quad
\overline{\mathfrak{b}}
\to\frac{{\rm e}^{\beta(\mu-h/2)}}{{\rm e}^{\beta(\mu+h/2)}+1},\quad
\mathfrak{c}\to\frac{{\rm e}^{2\beta\mu}}
                {{\rm e}^{\beta(\mu+h/2)}+{\rm e}^{\beta(\mu-h/2)}+1}.
\end{equation}
We observe that the Trotter-Suzuki number  $N$ enters only as a simple
parameter.  In the limit $N\to\infty$ we have 
\begin{equation*}
  \Phi^{(N\to\infty)}_{\mathfrak{b},{\overline{\mathfrak{b}}}}
=-2\pi\beta\Psi_{\mathfrak{b},{\overline{\mathfrak{b}}}}
  \quad\text{and}\quad
  \Phi^{(N\to\infty)}_\mathfrak{c}=-2\pi\beta\Psi_\mathfrak{c}.
\end{equation*}
It       turns   out      that    the    integration      contour  for
${\Psi_\mathfrak{c}}\ast\log\mathfrak{C}$  can  be moved  to  the real
axis.          Hence,      the     equation      for   $\mathfrak{b}$,
${\overline{\mathfrak{b}}}$ is taken  on the  line \mbox{$v=x\pm  {\rm
    i}\gamma$} and  $\mathfrak{c}$  on   the  real axis   \mbox{$v=x$}
leading to the set of equations (\ref{nle}) in section~\ref{sec:nle}. 

Lastly, we want to comment on the treatment  of the large $\mu$ limit.
In this  case the   Fourier coefficients  $\widehat{\mathfrak{c}}$ and
$\widehat{\mathfrak{C}}$ coincide      up  to   exponentially    small
corrections. Hence the last equation of (\ref{Fexp})  can be solved in
terms               of         $\widehat{{\mathfrak{B}}}$          and
$\widehat{\overline{\mathfrak{B}}}$ 
\begin{equation}
\begin{split}
  \widehat{\mathfrak{b}}&=
    {\rm sign}(k)\,\frac{1}{1+{\rm e}^{-k}}\,
    (\widehat{\Phi}_{+}-\widehat{\Phi}_{-})
    +\frac{1}{1+{\rm e}^{|k|}}\,(\widehat{{\mathfrak{B}}}
               -\widehat{\overline{\mathfrak{B}}}),\\
  \widehat{\overline{\mathfrak{b}}}&=
    {\rm sign}(k)\,\frac{1}{1+{\rm e}^{k}}\,
    (\widehat{\Phi}_{+}-\widehat{\Phi}_{-})
    -\frac{1}{1+{\rm e}^{|k|}}\,(\widehat{{\mathfrak{B}}}
               -\widehat{\overline{\mathfrak{B}}}),\\
  \widehat{\mathfrak{c}}=\widehat{\mathfrak{C}}&=
    {\rm sign}(k)\,\frac{1}{1+{\rm e}^{|k|}}\,
    (\widehat{\Phi}_{+}-\widehat{\Phi}_{-})
    -\frac{1}{1+{\rm e}^{k}}\,(\widehat{{\mathfrak{B}}}
               +{\rm e}^k\widehat{\overline{\mathfrak{B}}}).\\
\end{split}
\end{equation}
Applying the inverse Fourier transform and respecting the $v\to\infty$
limit 
\begin{equation}
\mathfrak{b}\to{\rm e}^{\beta h},\quad
\overline{\mathfrak{b}}\to{{\rm e}^{-\beta h}},\quad
\mathfrak{c}\to\frac{{\rm e}^{\beta\mu}}
{{\rm e}^{\beta h/2}+{\rm e}^{-\beta h/2}},
\end{equation}
we arrive at the integral equations (\ref{nle-hd}).
\end{appendix}


\end{document}